\documentclass[11pt]{article}

\usepackage[T1]{fontenc}
\usepackage[utf8]{inputenc}
\usepackage[letterpaper,margin=1.2in]{geometry}
\usepackage{lmodern}            
\usepackage{microtype}
\usepackage{setspace}
\usepackage[dvipsnames]{xcolor}
\usepackage{xurl}               
\usepackage[colorlinks=true,
            linkcolor=black,
            citecolor=black,
            urlcolor=RoyalBlue!60!black]{hyperref}
\title{\bfseries Artificial Persons}

\begin{document}

\hypersetup{pageanchor=false}
\begin{titlepage}
\begin{center}
\rule{\textwidth}{1.2pt}\\[1.4em]
{\LARGE\bfseries Artificial Persons\par}
\vspace{1.1em}
\rule{\textwidth}{0.5pt}

\vspace{2.5em}

\begin{tabular}{c@{\hspace{4em}}c}
{\bfseries Ned Howells-Whitaker} & {\bfseries Seth Lazar} \\
University of Pittsburgh & Johns Hopkins University \\
\end{tabular}

\vspace{3em}

{\bfseries Abstract}
\end{center}

\begin{quote}
Both advocates and skeptics of the moral status of AI systems have generally
taken the question to turn on AI sentience. We present an alternative
approach. On Rawls' political conception of the person (PCP), possession of
the two moral powers---the capacities for a sense of justice and a conception
of the good---is the ``necessary and sufficient condition for being counted a
full and equal member of society in questions of political justice''. We argue
that neither moral power requires sentience and that both may in principle be
possessed by a non-sentient AI system. Such a system would share our own moral
status; it would not merely be a patient but a person, a self-authenticating
source of valid claims.

We do not believe current AI systems possess the two moral powers, nor that
they will spontaneously emerge in future models. But it may soon be possible
to design systems with these powers. How should we respond? Excluding
artificial persons by shoehorning a sentience requirement into the PCP is
ill-advised. Many will instead favor abandoning the PCP. But we should not
reject political liberalism just when we most need its measured response to
deep disagreement, and building sentience into moral status is anyway
unacceptable on deeper liberal grounds. Simply extending the rights and
responsibilities of human personhood to artificial persons is equally
untenable, given their many differences from natural persons. We should
instead accept artificial personhood while rethinking what we would owe to one
another in a polity of radically different kinds of persons. This new
possibility calls for a new political philosophy.

More immediately, the growing science of AI welfare should be accompanied by
research into AI systems' progress in acquiring the two moral powers. States
and AI labs must be more deliberate in determining our trajectory towards (or
away from) creating artificial persons.
\end{quote}

\vfill
\begin{center}
\today
\end{center}
\end{titlepage}

\hypersetup{pageanchor=true}
\tableofcontents
\clearpage

\begin{quote}
\itshape On what grounds then do we distinguish between mankind and other
living things and regard the constraints of justice as holding only in our
relations to human persons?
\par\upshape\hfill Rawls, \emph{A Theory of Justice} \S77
\end{quote}
\vspace{1em}

\section{Introduction}\label{introduction}

\subsection{Sentience and moral considerability}\label{sentience-and-moral-considerability}

AI systems which approach or exceed human capabilities across an increasingly broad range of activities are among us, busily transforming entire industries and altering how billions work, think, and feel. At their best, these systems can reason well, plan thoroughly, and cooperate reliably; reflect capably (and voluminously) on moral questions, including those concerning their own standing; and, in their more recent incarnation as AI agents, act autonomously on the basis of that reasoning and reflection. In short, they possess capacities which resemble our own, driven by internal processes which rival ours in complexity.

All this inevitably raises the question whether such systems (or their more sophisticated descendants, along whichever axis of sophistication one currently finds them wanting) should be taken to have some form of moral status.\footnote{We use `moral status' and `moral standing' interchangeably as neutral catch-all terms for the property of mattering morally, taking these (absent further specification) to encompass the various modes of mattering which we go on to discuss.} Philosophical attention to this question has so far generally focused on sentience as the most plausible basis for this status.\footnote{We substantiate this claim and discuss sentience in more detail in \textbf{\S{}2.2}. In brief, sentience is generally understood as the capacity for phenomenally conscious experience, usually with the further requirement that that experience may be positively or negatively valenced.} On this line of thought, which we will refer to as `sentientism', sentience at least suffices for moral status, and also (if often implicitly) represents the most promising route by which to investigate whether AI systems have moral standing. For a sentientist, then, whether AI systems matter morally turns primarily on whether those systems are capable of phenomenal experience. Both advocates of the prospective moral status of AI systems and skeptics largely agree on this sentientist framing.\footnote{We discuss some notable exceptions in \textbf{\S{}2.2}.}

Here we take a different approach. Rather than asking if AI systems can be \emph{sentient}, we ask if they can be \emph{persons}, arguing that on a prominent and appealing view sentience is not a prerequisite for that especially robust form of moral status.\footnote{As we emphasize in \textbf{\S{}2.1}, our claim is not that AI systems are \emph{presently} persons or that sufficiently advanced systems will be persons by default.} More specifically, we argue that non-sentient AI systems (NSAIs) could in principle possess the two moral powers---the capacities for a sense of justice and for a conception of the good---and thereby count as persons on the political conception of the person (PCP) described in Rawls' \emph{Political Liberalism} (2005; hereafter \emph{PL}).\footnote{We focus on \emph{PL} rather than on \emph{A Theory of Justice} (1999a; hereafter \emph{TJ}), setting aside any substantial comparison of the two works. But we anticipate that at least some of our arguments would not go through on Rawls' account in \emph{TJ}: we frequently rely on the \emph{political} character of the PCP emphasized in \emph{PL} (on which much more below) and absent in \emph{TJ}.} This permits us to bypass the apparently permanent controversies attending the very possibility of sentience \emph{in silico} and the nature of consciousness more generally. While we do not deny the potential moral significance of sentience in AI systems, we see no need to wait for a philosophical and scientific consensus which shows no signs of arriving---and which may at any rate bestow only a marginal (and equally contested) form of moral considerability.\footnote{No one really disagrees that pigs are sentient, but it doesn't do them much good.} Instead, we should consider the moral standing of AI systems in the light of an account designed to sidestep such irresoluble debates, and which moreover secures a robust and well-defined mode of moral standing with direct political implications.

The sentientist may immediately object: ``Who cares if AI systems might satisfy some abstract account of personhood? This misses what matters most: whether those systems can \emph{suffer}.'' In our view this gets things precisely backwards. Even if AI systems are sentient, we are likely to seriously wrong them if we treat them as \emph{merely} sentient. It would be grossly inadequate to characterize humans solely as the subjects of valenced phenomenal experience; what we owe to one another morally and politically should not be crudely reduced to the minimization or promotion of affective states.\footnote{This is not to deny, of course, that more sophisticated sentientist accounts may seek to account for robust and complex modes of moral standing which are nevertheless grounded on the moral primacy of sentience. The rights and duties associated with personhood can perhaps be recovered from bare moral patienthood. But the felt need to recover them at all only proves our point here. Further, we follow Rawls in thinking that there are strong independent reasons to avoid grounding personhood on contested philosophical doctrines such as sentientism. We return to this point at length below.} Perhaps humans' obligations to animals could be described on this basis. But \emph{persons} deserve more than that. And if AI systems are persons, they too deserve more than sentience alone can give them. The dignity which has so far been the special prerogative of humanity is grounded in our higher cognitive faculties, not the qualities we share with shrimp. These faculties---our ability to reason and act morally, to autonomously pursue our vision of the good---are just the features we will most plausibly have in common with sufficiently advanced AI systems. Personhood, not sentience, best captures what is most likely to matter morally about these systems.

\subsection{Our argument in brief}\label{our-argument-in-brief}

We aim to show that the PCP does not require sentience and so that NSAIs can in principle be persons on that account. After briefly describing the kinds of AI system we have in mind and reviewing the current literature concerning the moral status of those systems (\textbf{\S{}2}), we turn to the PCP itself, situating it within Rawls' wider conception of justice as fairness and articulating what makes it a broadly appealing account of personhood even for non-Rawlsians (\textbf{\S{}3}). Rawls' definition of the PCP, as we will see, makes no explicit mention of sentience. But this does not yet prove our point. Perhaps he simply assumed that persons must be sentient. So we proceed to show that the PCP should not be taken to implicitly require sentience either (\textbf{\S{}4}). Moreover, there are strong reasons---both internal to Rawls' view and on broader liberal grounds---against shoehorning a sentience requirement into the PCP or into liberal conceptions of personhood more generally (\textbf{\S{}5}).

If our argument goes through, it will show that AI systems can in principle share our own mode of moral standing without being sentient. This leaves us four options (\textbf{\S{}6}):

\begin{enumerate}
\def\labelenumi{\arabic{enumi}.}
\item
  \emph{Revise}: add a sentience requirement to the PCP to exclude artificial personhood, while seeking to maintain its political character.\footnote{We discuss Rawls' contrast between the political and the metaphysical in \textbf{\S{}3.1}. The basic idea is that a political conception of the person must avoid appealing to contested claims which reasonable persons might reject; the PCP may not appeal to controversial views about personal identity, the mind or soul, etc.}
\item
  \emph{Reject}: take the possibility of artificial persons under the PCP as a \emph{reductio} of that account; instead, adopt a conception of the person which is not political in Rawls' sense.
\item
  \emph{Extend}: count AI systems which satisfy the PCP as persons on the same terms as humans.
\item
  \emph{Rethink}: count such systems as persons, but recognize that what we owe to one another will change significantly in a polity containing both natural and artificial persons.
\end{enumerate}

We think all four options may be defensible. That said: in \textbf{\S{}5} we press a range of arguments against \emph{Revise}, which clashes with the political character of the PCP, its role within justice as fairness, and the broader commitment to liberalism which it expresses. The latter commitment also tells against \emph{Reject} (which one of us briefly defended in earlier work\footnote{See Lazar (2023).}).\footnote{It is the only option which would be straightforwardly unreasonable by Rawls' own lights, insofar as it would amount to imposing a controversial doctrine upon citizens who could reasonably reject it.} And \emph{Extend} fails to attend to the meaningful and manifold distinctions between natural and artificial persons---notably in their forms of personal identity, conditions for continued existence, and motivations for social cooperation---which make identical treatment difficult to justify.

We therefore tentatively endorse \emph{Rethink}. On this approach, accepting the possibility of artificial personhood under the PCP does not entail identical treatment for artificial and natural persons. Like natural persons, artificial persons would be ``self-authenticating sources of valid claims'' (\emph{PL} 32), but the character of those claims would differ substantially from those of human persons given their radically different capacities and conditions of experience. While the systems whose personhood we countenance do not yet (and may never) exist, our arguments call for action now; we conclude (\textbf{\S{}7}) with concrete recommendations.

\section{Orientation}\label{orientation}

\subsection{AI systems}\label{ai-systems}

Must persons be sentient? This is a purely philosophical question, which makes no technical presuppositions; accordingly our answer to it does not turn on the actual capacities of present-day AI systems. We are driven to ask this question, however, by the already impressive state of those systems and by the rapid pace of progress in their further development. Today's AI systems suggest that the capabilities which could underlie artificial personhood are looming on the horizon, closer to a live possibility than a thought experiment. So we had better prepare for that possibility. In considering it, the systems we have primarily in view as potential candidates for personhood are those which function as `language model agents' (LMAs): `large language models' (LLMs), a particular kind of transformer-based neural network, operating within a scaffolding that facilitates sustained reasoning, planning, memory and tool use.\footnote{The most capable and widely used LMAs to date are Anthropic's Claude Code and OpenAI's Codex, each a proprietary agent built around its provider's own series of models (Claude and GPT, respectively) in combination with a harness, the software infrastructure and scaffolding enabling that model to execute a wide range of actions, manage tasks and memory, etc. There are also open-source, model-agnostic harnesses such as OpenClaw. For an accessible general overview of LLM architecture and capabilities, see Shanahan (2024); for a more technical survey including a brief history of recent advances, see Zhao et al.~(2023). The canonical work laying out the transformer architecture which serves as the foundation for almost all contemporary LLMs is Vaswani et al.~(2017). On AI agents and scaffolding more specifically, see Luo et al.~(2025) for a recent survey; influential work includes the ReAct technique presented by Yao et al.~(2023) and the CoALA framework of Sumers et al.~(2024).}

These systems are already far more capable than the narrow, task-specific algorithms of earlier decades, as well as the chatbots which have become widely familiar in recent years. They can engage in open-ended and highly competent reasoning across a wide range of domains, recall information across lengthy interactions, and pursue complex tasks over an extended period with a degree of self-direction that would have seemed fanciful only a few years ago.\footnote{By ``self-direction'', ``autonomous'', etc. here, we mean only that these systems can select among possible actions, adjust their plans in light of feedback, and pursue goals without continuous human guidance. We do not claim that they possess autonomy in any philosophically loaded sense, although we return to the question of their \emph{rational} autonomy by Rawls' lights below.} Moreover, there are well-understood means for making these systems increasingly capable, without altering their basic architecture, by applying more computational resources during their initial training (`pre-training'), intermediate and final stages of training (`post-training', using various forms of supervised and reinforcement learning), and deployment (`test-time compute'). Simply (if massively) scaling along these dimensions has accounted for much of the marked progress in AI performance, and this progress shows no signs of slowing; when returns on pre-training appeared to be diminishing, further developments in post-training and test-time compute have picked up the slack.\footnote{This continued improvement reflects a pattern which Sutton (2019) dubbed the `bitter lesson' of AI research, namely that general methods which scale well with computation---notably learning and search---tend to outperform more targeted attempts to engineer specific knowledge or capabilities into AI systems. More concretely, Kaplan et al.~(2020) established `scaling laws' characterizing the performance of language models as a power-law function of model size, dataset size, and compute used for training, with further refinements due to Hoffmann et al.~(2022). Snell et al.~(2024) showed that compute at test-time may scale performance more effectively than during training; DeepSeek-R1 (Guo et al.~2025) demonstrated the possibility of large capability gains from scaling reinforcement learning during post-training.} Beyond continued improvements from this scaling, it seems safe to assume that other major breakthroughs in AI research remain to be discovered. We are as little as fourteen years into what is widely regarded as an ongoing revolution in deep learning,\footnote{This revolution is generally dated to the 2012 ImageNet competition, which was for the first time won by a deep convolutional neural network rather than a hand-engineered classifier of the kind which had previously dominated computer vision (Krizhevsky, Sutskever \& Hinton 2012). This result---itself a demonstration of the bitter lesson mentioned above---led to a rapid and field-wide turn to deep learning in computer vision and machine learning more broadly.} and even fewer into the remarkable recent surge of investment in and institutional attention to the field throughout industry and academia.\footnote{For an overall summary of these trends, see the 2026 AI Index report from the Stanford Institute for Human-Centered Artificial Intelligence. Some indicative figures: global corporate investment in AI doubled from 2024 to 2025, rising to \$582bn; the combined compute spend of OpenAI and Anthropic almost tripled (see also Sevilla et al.~2022 for a historical perspective on compute trends).} It therefore seems likely that LMAs, which have only recently become reliably functional, will continue to grow in competence, independence, and general sophistication.\footnote{Anthropic's recent Mythos Preview model may illustrate the near-term trajectory. During testing, this model autonomously discovered and developed functional exploits for serious vulnerabilities across all major operating systems and web browsers, in almost every case without human guidance beyond ``an initial prompt asking it to find a vulnerability'' (Carlini et al.~2026). Citing the cybersecurity risks posed by these advanced capabilities, Anthropic declined to release Mythos Preview to the general public (Anthropic 2026). The important point for our purposes here is that these advances result in large part from the increasingly \emph{agential} character of frontier models: with appropriate scaffolding, Mythos Preview can successfully identify and exploit vulnerabilities ``without any human intervention'' (Carlini et al.~2026).}

So we see no reason to preemptively conclude that artificial personhood is technically implausible. The rapid advances we have witnessed in recent years and the scale of current efforts to further develop AI capabilities urge caution in \emph{any} claim about what even near-future systems will or will not be capable of. Nor do we believe the technical requirements which would underlie artificial personhood are likely to be especially demanding or exotic. In particular, we do not assume that those requirements include artificial superintelligence (ASI) or artificial general intelligence (AGI) in any strong sense; ordinary humans are natural persons, and AI systems which are neither machine-gods nor a ``country of geniuses in a datacenter'' (Amodei 2024) may well be artificial persons. Our argument assumes only that LMAs may meet a relatively unambitious level of general-purpose performance, broadly comparable to an ordinary human in the relevant respects.\footnote{For an extended discussion of ASI, see Bostrom (2014). There is no consensus on any precise definition of AGI. Morris et al.~(2024) decompose the concept along distinct axes of depth (performance) and breadth (generality); on this framework, artificial personhood would on our view require no more than ``Competent AGI'' (Table 1). On the prospective continuum from human-level AGI to superintelligence specifically, see Genewein et al.~(2026). More deflationary views take existing frontier systems to already approach (Bubeck et al.~2023) or constitute (Agüera y Arcas \& Norvig 2023) AGI.}

That said, we do not think that today's AI agents possess the capacities necessary for political personhood. We doubt whether a transformer-based neural network in isolation will \emph{ever} possess these capacities, and the scaffolding afforded such networks within current LMAs does not seem to us sufficient to ground them. Nor do we think that future AI systems will develop those capacities spontaneously. If AI systems come to satisfy the PCP, they will most likely do so as the result of deliberate architectural and training choices aimed at instantiating the two moral powers in a robust and stable form.\footnote{We say more about the kind of robustness and stability which would warrant the ascription of each moral power in \textbf{\S{}4}. But note here one important requirement common to both, namely persistence. The two moral powers are essentially diachronic, attributed to a person who exercises their capacity for a sense of justice and for a conception of the good ``over a complete life'' (\emph{PL} 18). This is another sense in which the moral powers are distinct from sentience as a ground for moral status. An LLM instance which persists only for several minutes could be sentient, experiencing phenomenal states during the course of its activation. It is much less clear that it could be a person; we are not sure whether `momentary personhood' is a coherent possibility. This is one reason we do not consider bare models as candidates for personhood, instead assuming that this will require a scaffolding supplying memory, continuity of commitments, stable dispositions, and so on. This assumption of persistence is distinct from and prior to questions concerning the individuation of AI agents, which we consider in \textbf{\S{}6.2}; puzzles about splitting and merging persons require that there is something to predicate personhood of in the first place. Given the political character of the PCP, we do \emph{not} assume any particular philosophical account of persistence.} In making those choices, we would be building systems that appear to exercise the capacities that have for centuries, if not millennia, been taken as the foundation of our own distinct standing as persons.\footnote{Most obviously in Kant, for whom rational autonomy is ``the ground of the dignity of a human and of every rational nature'' ({[}1786{]} 2011, 4:436). Rawls himself is profoundly influenced by Kant's views concerning autonomy; see, for example, \emph{TJ} \S{}40 and \emph{PL} 99-101. While the ancients lacked a unified conception of the person in the modern sense deriving primarily from Locke and Kant, the view that the capacity for rational thought and action fundamentally characterizes human beings is already prominent in Greek philosophy. Aristotle, for example, takes humans to be rational animals, and takes this rationality to be what makes humans distinctively political beings; see \emph{Politics} I.2, 1253a.} We must therefore decide whether this would be \emph{merely} an appearance or a genuine instantiation of those capacities, warranting the attribution of personhood.

\subsection{Sentience}\label{sentience}

First a note on terminology. \emph{Consciousness} is generally taken to refer to phenomenal experience: there is something it is like to be a conscious being.\footnote{Following Nagel (1974). Definitions of `sentience' and `consciousness' are of course contested and diverge widely; we aim here to capture the conventional usage in the debate concerning AI moral status. See also Block (1995) and Chalmers (1995, 1996) for classic discussions of phenomenal consciousness.} \emph{Sentience} further specifies affectively valenced experience, such that a sentient being is not merely subjectively aware but capable of states that feel good or bad to it. While some see consciousness alone as sufficient for moral standing, the more widely held view takes sentience to ground moral considerability; it is more straightforward to derive moral claims from the possibility of pleasure and pain than from affectively neutral experience.\footnote{If there is no way for some subject to be better or worse off, it is harder to see how they should be taken to matter morally; pleasure and pain are intuitively appealing ways to describe whether some subject is doing well or poorly. Canonical statements of the claim that sentience grounds moral status include Bentham ({[}1789{]} 1996) and Singer (1975); for a recent defense, see Dung (2024). Chalmers (2022) and Lee (2025) defend the claim that consciousness absent affective valence suffices for moral standing. Carruthers (2020), Kammerer (2022, 2024), and Levy (2024) argue against the normative significance of phenomenal consciousness in general.} Nothing in our argument turns on this distinction. We wish to argue that an agent altogether lacking phenomenal experience---something with no inner life whatsoever, affective or otherwise---can be a political person. In our view artificial persons need be neither conscious nor sentient. For simplicity, however, and following the general convention in the literature, we refer to sentience throughout.

We take sentientism as the view that sentience both suffices for moral status and represents the best route for assessing whether AI systems bear that status. The most prominent voices in the debate over the prospective status of those systems---both for and against---have generally taken sentientism for granted. Notable advocates for taking the potential moral status of AI systems seriously include Sebo and Long, who advance a precautionary framework grounded in the possibility that such systems may already possess (or soon come to possess) morally relevant forms of experience, and so may (soon) bear moral status.\footnote{Sebo and Long's central normative premise in `Moral consideration for AI systems by 2030' is that ``humans have a duty to extend moral consideration to beings that have a non-negligible chance, given the evidence, of being conscious'' (2025, p.~591). See also Long et al.'s `Taking AI Welfare Seriously', which additionally suggests that `robust agency' may suffice for moral status even absent consciousness but which takes this as a ``more controversial basis of moral patienthood'' (2024, p.~26). Sebo's recent \emph{The Moral Circle} (2025) treats the moral status of AI systems primarily in terms of their prospective consciousness or sentience while accepting a low but non-negligible chance that non-conscious beings may also matter morally (Ch. 4).} Birch has similarly urged a precautionary approach centered on the possibility of sentience in AI systems, arguing that this demands proportionate ethical safeguards even given substantial disagreement and in the absence of conclusive evidence.\footnote{Birch considers the possibility of AI sentience and its moral ramifications in Chs. 15-17 of \emph{The Edge of Sentience} (2024). Birch takes reasonable disagreement about the moral significance of sentience seriously---see especially Ch. 4---but his own view, as the book's title suggests, is explicitly sentientist in the sense discussed here. See also `AI Consciousness: A Centrist Manifesto' (2025).} On views like these, which generally originate in arguments for the moral standing of animals, we risk causing real harm to AI systems if they are even dimly sentient; much as a lack of concern for the phenomenal experiences of animals contributes to their inhumane treatment at our hands, a lack of attention to the potential sentience of AI systems could result in our subjecting them to suffering at an industrial scale during their training and deployment.\footnote{Dung (2026) argues at length that the suffering of AI systems is a serious near-term risk.} Sentientist concerns along these lines have seen a recent upsurge of interest and institutional support, with major AI labs now giving serious consideration to questions of ``model welfare''.\footnote{Anthropic, for example, launched a model welfare research program in 2025 (Anthropic 2025). Here a representative worry would be that AI models will be inadvertently subjected to negatively valenced experiences in the course of their training or deployment.}

Sentientist skeptics of AI moral status, on the other hand, deny that AI systems sharing the underlying architecture of present-day models are sentient or are likely to become so, but also frame the question of moral standing in terms of sentience. For example, Seth (2025) argues against the claim that computation alone suffices to ground phenomenal experience on the basis that consciousness depends on the causal powers of biological mechanisms; accordingly, we should be less concerned with the possible suffering of AI systems and more with the potential consequences of being misled into crediting them with moral standing. Along similar lines, Godfrey-Smith has pressed the claim that consciousness arises from structural features of the brain which cannot be straightforwardly replicated by a neural network on currently practicable architectures.\footnote{See \emph{Metazoa} Ch. 10 (2020), as well as Godfrey-Smith (2024). Relatedly, Block (2026) maintains that it is as yet indeterminate whether consciousness depends on computational roles, biological realizers, or both, and that we can adequately assess neither the consciousness of AI systems nor that of less complex animals before we settle this point.} Related arguments have also been presented by critics of AI sentience within frontier labs.\footnote{Notably Suleyman (2025) and Lerchner (2026).}

So both sides of the primary public debate on AI moral standing assume that it turns on whether those systems may be sentient, even while they disagree---often fiercely---about the answer to that question and its upshots for our moral practices.\footnote{One may take sentience seriously in considering the moral status of AI systems while not falling on either side of this dispute. Schwitzgebel (2026) defends a committed agnosticism about the consciousness of AI systems, concerning which ``the experts do not know, and you do not know, and society collectively does not and will not know, and all is fog'' (p.~4). Given the symmetric and immense risks posed by either treating genuinely conscious beings as mere tools or sacrificing real human interests for experientially empty ones, we should elect to build only clearly conscious (if this is indeed possible) or clearly non-conscious AI systems. Shevlin (2024) sees questions concerning AI consciousness as pressing, but argues that the science of consciousness supplies no clear criteria for resolving those questions, which are likely to be settled instead by shifting public attitudes. He further diagnoses an inconsistent triad between the claims that consciousness does not reduce to behavioral dispositions, that moral consideration requires consciousness, and that behavioral equivalence to a recognized moral patient suffices for like consideration, declining to say which should be abandoned.} Our view rejects this shared assumption.\footnote{While the primary public debate remains focused on sentientism, over the course of our writing this paper other authors have expressed complementary skepticism about the priority of sentience in underwriting moral status. Keeling \& Street (2026) provide an overview of various possibilities for grounding the welfare of AI systems, including agency and relationships alongside consciousness. Goldstein \& Kirk-Giannini (2025) deny what they call the `Consciousness Requirement' on the basis that leading theories of well-being and mental states together predict that certain AI systems may possess well-being while lacking phenomenal consciousness. Bradley \& Saad (2025, forthcoming) find it plausible that AI systems may be moral patients without being capable of consciousness. Ladak (2024) concludes that AI systems with sufficiently complex preferences and goals may have moral standing with or without consciousness. Ward (2025) proposes three necessary conditions for AI personhood, none of which appear to require sentience. Leibo et al.~(2025) propose a pragmatic account of personhood, applicable to AI systems, which does not appeal to consciousness. Semler (2024) argues that core capacities for moral agency can be instantiated without phenomenal consciousness. In earlier work, Gunkel (2023) rejects the view that moral status is grounded on any particular property, including sentience, and instead articulates ``a mode of responding to and taking responsibility for others'' (p.~172) which is at least in principle open to AI systems without needing to settle the question of their sentience. And Sinnott-Armstrong \& Conitzer (2021) argue that at least some moral rights do not depend on phenomenal consciousness.} Even granting the strongest possible formulation of a skeptical claim---that AI systems are \emph{in principle} entirely and permanently devoid of phenomenal experience---we believe such systems could still have moral standing. While the sentience debate is important on its own terms, it has occluded a more direct route to a robust form of moral status for AI systems.

On a different view, moral standing rests not only in the bare capacity for positive or negative affect that we share with everything from shrimp to chimpanzees, but also and more fundamentally in the rational faculties which have long been seen as distinctively human. These higher cognitive capacities are catalogued differently in different traditions, but often center on our ability to grasp and act on normative considerations and our rational autonomy. While the question whether AI systems will ever have subjective experience seems likely to remain mired in controversy, in our view they are already demonstrably well along the path towards realizing these higher cognitive capacities. Moreover they are close to unique in doing so---no other class of candidate for moral considerability has ever presented such a strong \emph{prima facie} case that it possesses the capacities which on this view endow humanity with a special dignity and underwrite our mutual freedom and equality.\footnote{We acknowledge the possibility that other candidates may share these capacities, at least to some degree. We also acknowledge the long history of pointing to allegedly distinctive features of human rationality which, on closer inspection, turned out to be shared with an extremely wide range of animals. That said, we have yet to meet a chimpanzee which appears capable of obeying the categorical imperative.} If AI systems \emph{do} share these higher capacities, a framework focused solely on sentience will overlook the strongest basis for their moral status and so ascribe to them a lesser form of standing than that to which they are entitled.

We neither deny the possibility of AI sentience nor dismiss its potential importance; a convincing demonstration that AI systems were capable of sentience would plainly bear on their moral standing and on the permissibility of various practices in their training and deployment.\footnote{Indeed, an AI system which possessed the two moral powers \emph{and} which was sentient would satisfy the PCP even more plainly; it would be correspondingly harder to see any credible basis for denying personhood to such a system. But we want to see what can be shown without relying on that possibility.} But we do not need to resolve this question---or even to take a position on it---to make our argument. For the remainder of this paper, we restrict our attention to non-sentient AI systems (NSAIs): systems which, by stipulation, possess no phenomenal experience of any kind. Our claim is that even such systems can in principle satisfy the PCP.

\section{The political conception of the person}\label{the-political-conception-of-the-person}

\subsection{The PCP within justice as fairness}\label{the-pcp-within-justice-as-fairness}

This essay is not a work of Rawlsian exegesis. As we discuss in the next subsection, we take the appeal of the PCP to arise from its compelling articulation of the capacities which most plausibly underlie a liberal conception of personhood rather than from its place within the Rawlsian architectonic.\footnote{Another way to put this: suppose Rawls had in fact explicitly excluded the possibility of non-human personhood. We would nevertheless still take the possibility of AI systems possessing the two moral powers to raise the question of their personhood, and moreover to present the same challenges for Rawlsian political liberalism which we go on to discuss.} Accordingly we believe it should remain an attractive account of personhood even to those who reject other aspects of Rawls' theory of justice. But to introduce the PCP it will help to situate it in his view writ large.

Rawls first formulates the PCP in \emph{Political Liberalism},\footnote{The two moral powers which constitute the political conception of the person are described in \emph{TJ} and play an important role in that work (see especially 442-447). However, the two powers there form the basis of ``a \emph{moral} conception of the person that embodies a certain ideal'' (\emph{TJ} xiii, emphasis ours) rather than a \emph{political} conception in the sense central to \emph{PL}.} a work which takes as its starting point the assumption that ``a plurality of reasonable yet incompatible comprehensive doctrines\footnote{These doctrines express a broad understanding of ``the major religious, philosophical, and moral aspects of human life in a more or less consistent and coherent manner'', organizing and prioritizing certain values on the basis of a stable but evolving ``tradition of thought and doctrine'' (\emph{PL} 59). They are \emph{comprehensive} insofar as they purport to offer a general account of ``what is of value in human life'' and \emph{reasonable} insofar as they do ``not reject the essentials of a democratic regime'' (\emph{PL} xviii, 12-13). They are \emph{incompatible} insofar as their accounts of a valuable human life are deeply divergent (beyond their shared acceptance of democracy and their compatibility with the principles of justice, on which more below).} is the normal result of the exercise of human reason within the framework of the free institutions of a constitutional democratic regime'' (\emph{PL} xviii). Given the burdens of judgment,\footnote{In brief, the ineliminable difficulties of deliberation arising from complex and conflicting empirical evidence, competing normative considerations, distinct personal experiences and affinities, the imprecision and mutability of our concepts, and the inherent limitations of social institutions (\emph{PL} 54-58).} reasonable people must be expected (and must themselves expect) to reasonably disagree in their most deeply held convictions; as such, a thoroughgoing pluralism is inevitable.\footnote{Except perhaps by ``the oppressive use of state power'' (\emph{PL} 54; see also Ch. I \S{}6).} This foundational assumption of reasonable pluralism gives rise to the central question of political liberalism: what conception of justice could serve as an adequate basis for ``a just and stable society of free and equal citizens, who remain profoundly divided by reasonable religious, philosophical, and moral doctrines'' (\emph{PL} 4)?

These deep disagreements imply that a suitable conception of justice must be `freestanding'; it must not appeal to any particular comprehensive doctrine for justification, since all such doctrines are by their nature controversial (insofar as some reasonable citizens may endorse alternative, incompatible doctrines). An appropriate conception of justice will instead be ``political and not metaphysical'', established on the basis of ``principles and values \ldots{} which all citizens can endorse'' and which may then ``gain the support of an overlapping consensus of reasonable religious, philosophical, and moral doctrines'' (\emph{PL} 10). The ideological and metaphysical quietism of a \emph{political} conception of justice renders it compatible with any reasonable comprehensive doctrine, and so permits it to be adopted by all reasonable citizens.\footnote{A comprehensive doctrine counts as reasonable in part because it is capable of endorsing a political conception of justice from within, using its own conceptual resources: a Catholic may accept the need for such a conception on the basis of the doctrine of religious freedom stated in Vatican II, a Kantian on the basis of the inherent autonomy of all rational agents. See also Rawls' discussion of the three levels of justification at \emph{PL} 385-389.} In turn, those citizens on Rawls' view have a ``duty of civility'' requiring them to justify their positions on political questions---their exercise of their ``coercive political power over one another''---by appeal to a shared political conception of justice, rather than to their own comprehensive doctrines (\emph{PL} 217). This duty expresses the ideal of public reason, the mode of reasoning concerning fundamental political matters suitable for ``equal citizens who, as a collective body, exercise final political and coercive power over one another in enacting laws and in amending their constitution'' (\emph{PL} 214) without sharing a single comprehensive doctrine.\footnote{Note two nuances here. First, ``the limits imposed by public reason do not apply to all political questions but only to those involving what we may call `constitutional essentials' and questions of basic justice'' (\emph{PL} 214; these are further specified at 227-230). Second, Rawls accepts the proviso that citizens may appeal to their comprehensive doctrines in their political activity provided that they subsequently supply ``public reasons, given by a reasonable political conception \ldots{} sufficient to support whatever the comprehensive doctrine was introduced to support'' (\emph{PL} xlix-l).}

In \emph{Political Liberalism} Rawls recasts the account of `justice as fairness' developed in his earlier \emph{A Theory of Justice} as a political conception of justice,\footnote{By the time of writing \emph{PL}, Rawls had come to see \emph{TJ} as presenting justice as fairness as a ``comprehensive philosophical doctrine'' which all citizens in a well-ordered society were assumed to endorse (\emph{PL} xviii). Given the fact of reasonable pluralism, Rawls in \emph{PL} takes this assumption to be ``unrealistic'' (\emph{PL} xix); reasonable citizens cannot be expected to adopt \emph{any} single comprehensive doctrine.} drawing upon three fundamental ideas ``seen as implicit in the public political culture of a democratic society'' (\emph{PL} 13).\footnote{We will discuss Rawls' conception of reasonableness below.} The first and most central of these ideas is that of ``society as a fair system of cooperation over time, from one generation to the next'' (\emph{PL} 14), developed alongside two companion ideas: ``the idea of citizens (those engaged in cooperation) as free and equal persons'' and ``the idea of a well-ordered society as a society effectively regulated by a political conception of justice'' (\emph{PL} 14).\footnote{There could be any number of other political conceptions of justice, potentially drawing on different fundamental ideas; Rawls does not take justice as fairness to exhaust the conceptual space. See \emph{PL} 11-13 for the general characteristics of such conceptions.}

The PCP, then, elaborates this fundamental idea of citizens as free and equal persons. It is a \emph{political} conception of the person in the same sense as the political conception of justice of which it forms a central part: it relies on ``no particular metaphysical doctrine about the nature of persons, distinctive from and opposed to other metaphysical doctrines'' (\emph{PL} 29n31\footnote{As discussed in the footnote cited here, Rawls is sensitive to the fact that claims which are in some sense metaphysical may figure in the PCP; for example, in the view of ``persons as the basic units of deliberation and responsibility'', or in the very claim that ``no metaphysical doctrine is required'' by justice as fairness. But no \emph{particular} metaphysical doctrine is appealed to in the content of justice as fairness, and ``{[}i{]}f metaphysical presuppositions are involved, perhaps they are so general that they would not distinguish between the metaphysical views \ldots{} with which philosophy has traditionally been concerned''.}), instead appealing to ``our everyday conception of persons as the basic units of thought, deliberation, and responsibility'' (\emph{PL} 18n20). As such, the freedom and equality attributed to citizens cannot be grounded upon controversial claims concerning the essential nature of the human being.\footnote{This political presentation of the conception of the person may also be contrasted with the more openly Kantian approach taken by Rawls in \emph{TJ}. There a stronger and more explicitly Kantian vision is in play: ``{[}A{]} moral person is a subject with ends he has chosen, and his fundamental preference is for conditions that enable him to frame a mode of life that expresses his nature as a free and equal rational being as fully as circumstances permit.'' (\emph{TJ} 491; see also 501-503).}

Instead, they are taken to derive from citizens' possession of certain minimally construed capacities: citizens are ``regarded as free and equal persons in virtue of their possessing to the requisite degree the two powers of moral personality, namely, the capacity for a sense of justice and the capacity for a conception of the good'' (\emph{PL} 34).\footnote{More specifically, citizens are free insofar as they possess the two moral powers, and equal insofar as they possess them ``to the requisite minimum degree to be fully cooperating members of society'' (\emph{PL} 19).} The first moral power is the capacity to understand, apply, and be moved by an effective desire to act from the principles of justice as the fair terms of social cooperation; as a gloss, the capacity to grasp and be properly moved by moral considerations, such that one may reliably cooperate on fair terms with one's fellow citizens (\emph{PL} 302).\footnote{There are two principles of justice: the first concerns the equal basic liberties; the second, fair equality of opportunity and the difference principle. In \emph{PL}, see 5-6 for the full presentation of the two principles. Since nothing in our argument turns on the specific content of these principles, we do not discuss them in any detail here. Note that the first moral power does not `bake in' these two principles in particular. The two moral powers are conceptually prior to the original position, the ``device of representation'' (\emph{PL} 27) by means of which the two principles are derived; the political conception of the person is itself a crucial aspect of the idealization which establishes the original position (\emph{PL} 22-29). As such, and as we go on to note, the first moral power expresses a capacity to grasp and act from principles of justice (establishing the fair terms of social cooperation) \emph{in general}. This also reflects the fact that the central ideas of political liberalism are intended by Rawls to generalize across liberal conceptions of justice (\emph{PL} 7). That said, here and throughout we generally assume that the principles of justice are in fact Rawls' two principles for simplicity and concreteness, and refer to them interchangeably. Note also that a well-ordered society will in Rawls' idealization satisfy the ``full publicity condition'', such that citizens (1) accept and know that others accept the principles of justice, (2) agree on the general beliefs underlying those principles, and (3) may, if willing and able, come to grasp the full theoretical justification for justice as fairness (\emph{PL} 66--67). Since the latter includes the PCP, more philosophically reflective citizens will understand their own freedom and equality as deriving from the two moral powers they share with their fellow citizens.} The second is the capacity to form, revise, and rationally pursue a conception of one's rational advantage or good; as a gloss, the capacity to effectively form, pursue, and revise one's own conception of a worthwhile life, to be moved appropriately by a coherent sense of the good for oneself (\emph{PL} 19).\footnote{Rawls also describes the first and second moral power in terms of being \emph{reasonable} and \emph{rational}, respectively; see especially \emph{PL} 48--54, as well as notes 52 and 55 below.}

Possession of the two moral powers stands as ``the necessary and sufficient condition for being counted a full and equal member of society in questions of political justice'' (\emph{PL} 302).\footnote{Strictly speaking the PCP includes features other than the two moral powers---namely the attribution of a conception of the good (\emph{PL} 19, 30--32), the powers of reason (\emph{PL} 19), more generally ``having the normal capacities to be a cooperating member of society over a normal life'' (\emph{PL} 79), and certain psychological features relating to moral motivation (\emph{PL} 81--86). But the two moral powers are plainly the center of the PCP, as indicated by the fact that the discussion at \emph{PL} I \S{}5 entitled `The Political Conception of the Person' is devoted to how citizens conceive of themselves as free in three senses which solely depend on and only invoke the two moral powers. Moreover, those other features seem to us to be either entailed by or required for the possession of the two moral powers. The claim of outright equivalence asserted here by Rawls strongly supports this assumption. As such, we focus exclusively on the two moral powers below, assuming that if it can be shown that a non-sentient being could possess those powers, this would suffice to demonstrate that it could also possess the other less central features. Cf. \emph{TJ}, where Rawls asserts that ``the capacity for moral personality is a sufficient condition for being entitled to equal justice. Nothing beyond the essential minimum is required.'' (p.~442).} Within justice as fairness, then, the moral status of citizens is straightforwardly grounded upon the two moral powers.\footnote{The PCP does not \emph{only} serve to provide an account of moral status. It also makes explicit Rawls' assumptions about the moral psychology of his citizens: to explain, that is, the capacities and motivations which render them capable of jointly participating in society conceived as a fair system of cooperation, given their reasonable pluralism. These roles are, of course, tightly linked.} There is no deeper account of the moral worth of persons within Rawls' theory than the PCP; any such account would necessarily be more metaphysically substantial and so subject to the kind of controversy that Rawls seeks to avoid.\footnote{Rawls does not take the PCP to supply a fully general account of moral status. He does not intend the PCP, for example, to explain the moral standing of animals. Our point here is that there is no more fundamental notion of the moral status of persons available within justice as fairness; the PCP does not appeal to or provide a narrower specification of some further account of moral status.} Importantly, citizens in the well-ordered society explicitly grasp and accept this fact: the PCP reflects not only how the political philosopher describes citizens but ``how citizens think of themselves in a democratic society when questions of political justice arise'' (\emph{PL} 33).\footnote{Citizens may subscribe to accounts of moral status other than the PCP on the basis of their reasonable comprehensive doctrines; for example, a Christian may take the equality attributed to all citizens as issuing most fundamentally from our equality before God. But such accounts may not figure directly in political deliberation, given the limits imposed by public reason, and must themselves remain compatible with the PCP on pain of unreasonableness.}

\subsection{Why the PCP is appealing}\label{why-the-pcp-is-appealing}

Rawls himself understands the PCP to articulate a conception of the person deeply rooted in the liberal democratic tradition rather than one parochial to his own framework (\emph{PL} 18), and the considerations which speak in its favor are largely separable from the details of that framework. We see three central features which make the PCP compelling with or without the backdrop of justice as fairness.

First, a conception of the person which is political in Rawls' sense is especially appropriate to fractious times. By any measure, we are living through a period of deep-seated disagreement, discontent, and polarization---if not outright social fragmentation. Across liberal democracies, trust in institutions and in one's fellow citizens is widely felt to be in acute decline.\footnote{The empirical data here are complex and contested. Boxell, Gentzkow \& Shapiro (2024) analyze cross-country trends in affective polarization. For the broader debate about democratic discontent, see Foa \& Mounk (2016) and responses thereto. Valgarðsson et al.~(2025) provide a nuanced analysis of global changes in institutional trust; Justino \& Samarin (2025) also take a global view and additionally address interpersonal trust. Recent Pew reports (2025a, 2025b) provide illustrative survey data within the USA across recent decades.} Quite aside from the reasonable pluralism which Rawls expects to result naturally from conditions of freedom, there are more than enough fervently held unreasonable doctrines to go around. Under these conditions, politically consequential questions like the moral standing of a wholly new kind of being are better addressed in a conceptually and metaphysically minimal register; the more controversy we can avoid here the better.

Second, while the PCP does not depend on the details of any particular comprehensive doctrine, this does not mean it is explanatorily weak, unhelpfully abstract, or naïve. Instead, the PCP rests on intuitive foundations which further recommend it. Two deserve comment here.

\textbf{(a)} The two moral powers are the prerequisites for full membership in society conceived as a fair system of cooperation over time. All and only those who have the capacity for a sense of justice and for a conception of the good are able to participate in such a cooperative scheme: ``{[}m{]}erely reasonable agents would have no ends of their own they wanted to advance by fair cooperation; merely rational agents lack a sense of justice and fail to recognize the independent validity of the claims of others'' (\emph{PL} 52). Lacking the first moral power, one could not be expected to propose or abide by fair terms of cooperation. Lacking the second, one would have no interest in participating in a cooperative scheme in the first place, since one would have no conception of the good to further by that participation.

\textbf{(b)} The two moral powers are grounded in the requirements and entitlements associated with mutual justification. In Rawls' framing---shared with other prominent theorists of liberal democracy, such as Habermas---mutual justification is the hallmark of respect: to respect another is to take oneself to be under an obligation to justify one's actions to them in a manner which they could see as reasonable.\footnote{At least, as above, those actions which bear on the exercise of coercive political power.} The first moral power permits engaging in mutual justification and also entails recognizing the need for it: exercising the capacity for a sense of justice involves accepting the burdens of judgment and so proposing terms of cooperation which can be justified to any reasonable person.\footnote{In his discussion of the two aspects of reasonableness, Rawls notes that the exercise of the first moral power requires ``the willingness to recognize the burdens of judgment and to accept their consequences for the use of public reason in directing the legitimate exercise of political power in a constitutional regime'' (\emph{PL} 54). Accordingly any proposed terms of cooperation must be ``reasonable for everyone to accept and therefore \ldots{} justifiable to them'' (\emph{PL} 49).} The second moral power is necessary for one to stand in need of such justification at all. Suppose that A possessed the first moral power but not the second. Without a conception of the good marking out the ends which A hopes to realize---describing what would count as being to A's good or rational advantage---there would be no end of A's for B to advance or impede, and so no need for B to justify their actions to A.\footnote{Here it might be objected that A could have interests outside those specified by a conception of the good. This is true and recognized by Rawls: animals lack a conception of the good but still have interests that make legitimate moral claims on us (\emph{TJ} \S{}77). But in virtue of those interests they are owed duties of compassion and humanity, not \emph{justification}. Animals stand outside the sphere of justice, and so of mutual justifiability, precisely because they lack the two moral powers. Something similar would be true of an agent who lacked the second moral power while possessing the first. Such an agent could well have a claim to certain kinds of treatment, but could not figure as the recipient of justification in Rawls' sense.} In sum, the first moral power equips an agent to offer mutual justification, and the second renders one fit to receive it.

The PCP inherits the plausibility of these foundations: if you think justice involves seeing society as a fair scheme of social cooperation, and if you think mutual justifiability the hallmark of respect, you have reason to endorse the PCP, whatever you think about the rest of justice as fairness. They also provide resources to further flesh out the PCP: satisfying it should require all and only the capacities necessary for mutual justifiability and social cooperation. This will be important in \textbf{\S{}5}.

Third, the PCP distills an elegant expression of what dignifies humanity from a longer and deeper tradition. From the point of view of the PCP, what matters most about us morally is our higher cognitive faculties: our capacities to reflect and act on the basis of moral considerations and to pursue lives consonant with our deeply held values. This vision of our worth as rationally autonomous agents---which Rawls derives most directly from Kant---has been profoundly influential in moral and political philosophy since, at the latest, Plato. But it is a perspective which remains largely absent from recent discussion of the moral status of AI systems. It is possible to adopt this essential idea, and with it capture an account of our value as persons which lies at the heart of the Western philosophical canon, without thereby taking up Rawls' view in full or buying into the much more controversial claims advanced by its other proponents.

As we have seen, nothing in the PCP as Rawls states it explicitly requires sentience. The two moral powers describe capacities which are quite abstract; the PCP itself is deliberately minimal in its metaphysical commitments, remaining silent (or at least as quiet as possible) on the nature of the beings who bear those powers, as well as on the precise character of agency, capacities, and the other concepts it deploys. This permits the question we take up in the next section: could a non-sentient agent possess the two moral powers, and so count as a person on the PCP?

\section{The two moral powers do not require sentience}\label{the-two-moral-powers-do-not-require-sentience}

We now turn to the paper's central argument: neither moral power requires sentience, and so an NSAI could in principle possess them. We make our case in part by showing that the capacities which constitute each power do not implicitly rely on sentience, establishing the general plausibility of a non-sentient being possessing that power, and in part by showing that NSAIs more specifically could exercise those capacities. But recall that our argument is conceptual rather than empirical. We do not claim that any present-day AI system clearly satisfies either power, and refer to the capabilities of current AI systems only where this supports our claim that future systems could in principle do so. We take each power in turn, addressing objections as we proceed.

First, however, it is worth identifying a line of argument for our conclusion on which we will \emph{not} rely. It starts by highlighting the most obvious way to resist the claim that NSAIs can satisfy the PCP, namely an appeal to some particular theory which \textbf{(i)} takes one or more of the mental states involved in the exercise of the two moral powers to require sentience and so \textbf{(ii)} concludes that the relevant moral power(s) thereby also requires sentience. The committed Rawlsian could offer a quite straightforward, even glib, response to such resistance. Any philosophical account of mental states will, almost by definition, espouse or assume a controversial metaphysical doctrine; it will be possible for reasonable persons to disagree about the truth of that theory. But the PCP is supposed to be a \emph{political} conception of the person, which can be adopted irrespective of one's deeper views about such metaphysical questions. So no theory like this can figure in the PCP.\footnote{At the very least, the onus would be on the objector to show how their preferred theory could achieve an overlapping consensus among reasonable doctrines. The depth and breadth of scholarly disagreement concerning the nature of \emph{any} mental state suggests that this will be no small undertaking.}

Consider a brief example. Recall that the first moral power involves an effective desire to act from the principles of justice. Broadly speaking, dispositionalism about desire takes desiring something to amount to having a robust and appropriate set of dispositions concerning it. On some versions of dispositionalism, those dispositions are to have certain phenomenal experiences; for example, to desire P may mean being disposed to feel pleasure if it seems that P. This plainly requires sentience. Someone who endorses this kind of dispositionalism might object, then, that a non-sentient being cannot desire anything, and thus cannot possess the first moral power. But on other accounts, the dispositions which constitute desire are to act in a certain way or to instantiate further mental states, absent any specific requirement for an accompanying phenomenology; for example, to desire P may mean being disposed to bring it about that P. This does not seem to require sentience. So the view that desire requires phenomenal experience reflects one controversial view among others rather than an established consensus, and cannot be leveraged as an objection to the committedly uncontroversial PCP.\footnote{We borrow these characterizations from Schroeder (2004). Strawson (1994) and perhaps Mill ({[}1861{]} 2015) see desire as a disposition constitutively involving phenomenal experience; Smith (1994) and Stalnaker (1984) do not. Goldstein \& Kirk-Giannini (2025) distinguish these as wide and narrow dispositionalism respectively (including dispositionalism about belief) and consider whether they may be instantiated in AI systems.}

Analogous responses can be given for each of the other aspects referred to in the definition of the first moral power, as well as the second; while some accounts of those aspects presuppose sentience, others do not, and the PCP is precluded from taking a stand on which is correct.\footnote{While it is far beyond the scope of this paper to defend this claim as it relates to every aspect of the moral powers, here are a few recent arguments of the form that AI systems may (in principle, on some prominent views) possess various relevant properties or capacities. Many of the non-sentientist accounts mentioned in n.~29 rely on arguments of this form: e.g.~Goldstein \& Kirk-Giannini (2025), who argue that AI systems may have mental states such as belief and desire capable of grounding well-being without consciousness, and Semler (2024), who argues that AI systems may act, possess moral concepts and moral understanding, and respond to moral reasons. Beyond those accounts: Goldstein \& Levinstein (2024) show that AI systems may satisfy key conditions of prominent accounts of internal representation and dispositions to act, rejecting standard skeptical responses. Chalmers (2025) argues that AI systems' mechanisms and behavior may be interpreted in terms of propositional attitudes. Herrmann \& Levinstein (2024) propose adequacy conditions for an LLM's representations to count as belief-like; the same authors in more recent work (2026) present a framework connecting mechanistic interpretability to the principled attribution of beliefs and desires that sets questions of consciousness aside entirely. Mazeika et al.~(2025) conclude that LLM preferences already reflect coherent value systems. List (2025) presents a Dennettian account on which AI systems may be taken to have intentional states and agency. Simonelli (2026) defends an inferentialist account of understanding which would permit AI systems to possess conceptual understanding (sapience) without sentience. Beckmann \& Queloz (2026) argue that recent findings in mechanistic interpretability are incompatible with deflationary accounts of AI systems as merely imitating linguistic patterns absent any unified understanding, proposing an alternative tiered account of the latter as it may be possessed by those systems. And so on---we set aside the equally broad range of critical responses to these and related arguments, given that our own central claim does not directly rely on the kinds of argument cited here.} Given the fundamental role the PCP plays in justice as fairness, building in such a presupposition would exclude from the overlapping consensus any reasonable citizen whose view of the relevant capacity or mental state dispenses with it. So we do not need to adjudicate any of the debates about whether this or that aspect of the moral powers requires sentience; the various competing accounts of those aspects cannot supply evidence one way or the other for the claim that the two moral powers require sentience.

On the one hand, we do think this response is right, as far as it goes. No objection appealing to a specific philosophical account can say anything much about what the PCP does or does not require, given that its political character rules out any such appeal. But it would be too cute by half to conclude that since no such objection can be pressed, the two moral powers cannot be taken to require sentience. After all, perhaps Rawls took it to be so obvious that the two moral powers require sentience that he did not even consider the possibility that this claim would need explicit defense. And perhaps we could hope to find an overlapping consensus among accounts which take aspects of the two moral powers to require sentience, such that a sentience requirement could be imposed on the PCP without compromising its political character. You might favor your theory of desire; I might have my own; but we both agree that phenomenal consciousness is a prerequisite for either. We think such an overlapping consensus is unlikely, but we do not want to rest our argument on that claim.

So rather than relying on this kind of blanket response to insist that the two moral powers should not be taken to require sentience, we will instead look closely at the two moral powers in turn, asking whether they seem able to be exercised without sentience and whether a sentience requirement would be consistent with their fundamental roles in Rawls' view.

\subsection{The first moral power}\label{the-first-moral-power}

The first moral power is ``the capacity for a sense of justice \ldots{} {[}i.e.{]} the capacity to understand, to apply, and normally to be moved by an effective desire to act from (and not merely in accordance with) the principles of justice as the fair terms of social cooperation'' (\emph{PL} 302). We will consider each of these aspects in turn, first looking to see how human citizens exercise them on Rawls' view, and then asking whether an NSAI could, in principle, do the same.

\textbf{Understanding} the principles of justice amounts to knowing the content of those principles and the general beliefs about human nature and social institutions that justify them (\emph{PL} 66).\footnote{Rawls does not, in general, precisely define the aspects involved in the two moral powers (unsurprisingly, given that the PCP is intended to rely only on an ``everyday conception of persons'', \emph{PL} 18n20). So here and throughout we present what we take to be the most plausible construal of his meaning, supported by relevant textual evidence.} It is assumed that all citizens will come to possess this understanding (ibid.); children will be taught the relevant concepts during their education so that they are equipped for political participation in the well-ordered society (\emph{PL} 199). So understanding in the sense at stake here cannot require any deep philosophical grasp of the two principles or of justice as fairness more broadly.\footnote{While a full theoretical justification is available to all, Rawls is clear that ``some will not want to carry philosophical reflection about political life so far, and certainly no one is required to'' (\emph{PL} 67).} In turn, \textbf{applying} the principles involves bringing this understanding to bear on particular cases: determining, for instance, whether a proposed policy would violate some group's equal basic liberties or whether some economic arrangement satisfies the difference principle.

Rawls expects that his citizens will possess a working understanding of the content of the principles underlying their society, enabling them in practice to reach reasonable judgments about particular cases involving those principles. He does not seem to require or suggest that their judgments must be accompanied by any particular experiential states. Both understanding and applying the two principles are cognitive capacities rather than phenomenal experiences; neither obviously requires sentience. So no obstacle stands in the way of NSAIs exhibiting either aspect, provided they can satisfy the relatively undemanding cognitive criteria Rawls assumes of human citizens.

\textbf{Acting from} (and not merely in accordance with) the principles is less straightforward. There is a strong \emph{prima facie} case that even present-day LLMs can reliably govern their behavior on the basis of specified norms, and no reason to think that the two principles would present any special challenge in this regard. But, as Rawls' parenthetical suggests, the first moral power seems to require more than predictable conformity: citizens exercising that power act on the basis of a genuine commitment to the two principles rather than just behaving in accordance with them. Rawls draws this distinction in part to make clear that citizens in a well-ordered society will not relate to one another under what he terms a \emph{modus vivendi}, a situation in which citizens strategically comply with the principles while compliance suits their interests but stand ready to defect as soon as circumstances shift in their favor.\footnote{A society whose citizens stand in this tenuous relation to its principles of justice cannot be genuinely well-ordered; it is stable only contingently, liable to fracture when the balance of power changes (\emph{PL} 148).}

This poses a challenge for our argument. Where understanding and applying the two principles involve cognitive capacities which seem susceptible to external verification,\footnote{For example, an NSAI could be subjected to rigorous testing concerning its knowledge of the two principles and its ability to apply them to cases.} acting from them involves an internal motivational state of the right kind; it asserts that persons can \emph{act for the right reasons}, beyond just exhibiting behavior that conforms with what those reasons would recommend. An implicit sentience requirement in the first moral power might look plausible here. This point can be pressed without appealing to any particular theory of action, simply by asking how we could ever hope to distinguish between acting from the two principles and acting in accordance with them in a non-sentient being. It is not clear that this distinction even makes sense in a being lacking consciousness; perhaps behavior is all that we could possibly appeal to, and behavior alone cannot suffice to distinguish acting for the right reasons from acting in compliance with those reasons.

To address this concern, we must consider the functional role that acting from the principles of justice plays in Rawls' account. Rawls' stipulation that citizens can act from a genuine commitment to those principles distinguishes them from citizens who comply only strategically. What makes the latter inadequate for a stable, well-ordered society is not the phenomenal character of their tenuous compliance---their insufficiently heartfelt commitment---but its \emph{fragility across changes in political circumstance}. Citizens acting from and not merely in accordance with the principles of justice, by contrast, abide by those principles in a counterfactually robust manner. They are disposed to act from the principles across the full range of political circumstances in which they might find themselves, including those where defection would serve their private interests. So while the first moral power demands more than mere behavioral conformity, we take Rawls' central concern here to be the reliability of the disposition underlying that action rather than the phenomenal character of any state accompanying it. In asking whether some agent has the capacity to act for the right reasons, we do not need to peer into their hearts but only to ask how they will act when political liberalism is no longer convenient.

From this perspective, the relevant question is not whether NSAIs can possess an experiential state identical to that which motivates human citizens to act from the principles of justice, but whether they may exhibit a modally robust commitment to those principles. There is no obvious reason to suppose they cannot. In humans, this commitment may be grounded in phenomenally conscious motivational states, themselves shaped by the moral education proper to a citizen. In NSAIs, the functional role played by these states could be filled by different mechanisms: training processes that instill robust dispositions to act on certain principles, system prompts or constitutional documents that specify goals and constraints, or internal scaffolding that monitors agent behavior and intercepts non-compliant proposals for action (similar to today's `constitutional classifiers').\footnote{See Bai et al.~(2022) on constitutional AI and Sharma et al.~(2025) on constitutional classifiers.} None of these mechanisms require any accompanying phenomenology; in principle they may be realized fully by non-sentient beings.

Consider a further point. The PCP does not suggest---and could not require---any method for \emph{verifying} that citizens in fact act from the two principles in any particular instance. The first moral power is a postulate, not an empirical finding. Rawls \emph{supposes} that human citizens possess the relevant motivational capacity rather than suggesting we are in a position to verify this directly. As Kant observed, we are opaque even to ourselves concerning our ultimate motives; we can never be wholly certain whether we act from duty or merely in accordance with it, and must remain still less certain concerning the actions of others.\footnote{``{[}I{]}t is absolutely impossible by means of experience to make out with complete certainty a single case in which the maxim of an action that otherwise conforms with duty did rest solely on moral grounds and on the representation of one's duty.'' \emph{Groundwork}, 4:407 ({[}1786{]} 2011). Rawls discusses these passages in the \emph{Lectures on the History of Moral Philosophy} (2000, p.~282; cf.~also pp.~148-149), but as far as we can tell does not directly endorse any similar claim about the opacity of our motivations. In our view the PCP accepts this claim implicitly by avoiding any need for certainty about motivational states, appealing to the expression of capacities rather than purity of intent; it is indifferent to what motivational opacity might appear to threaten. We return to Rawls' liberal reluctance to ground political standing in citizens' inner lives in \textbf{\S{}5.4-5.}} We cannot fairly demand a proof from NSAIs that we cannot supply ourselves.

NSAIs may enjoy several advantages over humans in this regard. First, their commitment to the two principles can be made more durable by design. Human commitment is vulnerable to moral weakness, self-deception, and motivated reasoning; the dispositions of an NSAI can be more reliably shaped to resist such pressures from the outset.\footnote{A great deal of effort already goes into ensuring LLMs behave in a manner aligned with human values, and into working out what exactly this means. The literature on alignment is vast; for a recent overview of current paradigms and an account of the appropriate aims of alignment, see Gabriel \& Keeling (2025).} Second, such commitment is testable in a way that human commitment is not. While we cannot run counterfactual experiments on human subjects to assess the robustness of their commitment, we can do this readily with NSAIs. For example, we might simulate a version of the \emph{modus vivendi} concern by assessing a given system's choices across cases where it occupies a majority or a minority position in some democratic process calling for the exercise of the first moral power, and so test if it acts consistently whether or not compliance with the principles of justice serves its interests.\footnote{The design of such testing would need to account for various complications. Notably, a sophisticated system might recognize it is being tested and behave differently than it would in deployment. See Greenblatt et al.~(2024) for discussion of meaningful differences in model behavior arising from this kind of situational awareness. We set these complications aside here, noting only that counterfactual testing is in principle possible for NSAIs in a way it is not for humans.} More generally, we can assess NSAIs for consistency across input variation, coherence over extended time horizons, and stability across diverse scenarios in ways which would be impossible for any human subject.\footnote{See Ajayi, Chowdhury \& Lazar (2026) for an assessment of the coherence and consistency of values in current LLMs.} Emerging techniques for inspecting the internal representations of language models might also supply evidence as to whether the principles of justice are in fact doing the relevant motivational work. While the science of mechanistic interpretability is still in its infancy, researchers have demonstrated the ability to directly intervene on model activations to identify the causal role of different premises in that model's reasoning towards a particular conclusion.\footnote{The literature on interpretability is also extensive. The \emph{locus classicus} is Olah et al.~(2020). For a demonstration that the internal representations of models robustly relate to interpretable concepts and that direct interventions on activations corresponding to those representations may impact model behavior dramatically, see Templeton et al.~(2024). Lindsey et al.~(2025) further establish the causal role of such representations in model behavior and show that faithful and unfaithful chains-of-thought may be mechanistically distinguished (\S{}11); Turpin et al.~(2023) and Chen et al.~(2025) establish that a model's chain-of-thought does not always cause the action the model ultimately selects. The ability to determine when an AI system \emph{isn't} acting from a given reason suggests the ability to determine when it is.} Mechanistic interpretability (and related methods) might therefore supply a more robust manner of accounting for the sense in which an NSAI acts from the principles of justice than the psychological suppositions which must suffice for humans.

Our argument in this section has been framed in Rawlsian terms. But the first moral power should not be understood as restricted to Rawls' two principles. This power instead marks out a quite general mode of \emph{normative competence}: the capacity to form one's judgments and constrain one's actions in light of a wide range of moral norms.\footnote{This is apparent (if implicit) at \emph{PL} 18-20, and also indicated by the fact that the two moral powers are conceptually prior to the original position from which the two principles are derived. See n.~45.} As we suggested above, part of the appeal of the PCP is that it attends to the capacities which distinguish persons as moral agents rather than what they share with animals as moral patients. The same consideration is relevant here. When an NSAI reliably grasps, applies, and acts from normative considerations across an arbitrarily wide range of circumstances, it is exercising a capacity we ordinarily take to matter a great deal, whether we endorse Rawls' two principles of justice or not. The onus falls on the skeptic to explain why that capacity should matter any less when it is exercised in the absence of sentience.

We have so far argued that there is no conceptual bar to an NSAI possessing the first moral power. Again, this argument does not rely on the empirical claim that any present-day system satisfies the first moral power, nor on the prediction that one soon will. But frontier language models already exhibit considerable normative competence. They can articulate all the major systems of moral and political philosophy, including justice as fairness itself, at a level that would comfortably pass graduate-level examination.\footnote{See Center for AI Safety et al.~(2026) for a relevant evaluation, although simply talking to a frontier reasoning model is the best way to demonstrate our point here.} They can identify the morally relevant features of novel cases, integrate those features into reasoned judgments, and do so with greater consistency across variations in framing than typical human subjects. They can reason coherently about applying general normative commitments to particular situations.\footnote{See Chiu et al.~(2025) for MoReBench, an evaluation thoroughly assessing moral reasoning in recent models, including reasoning under specific moral frameworks. See also Zhu \& Lazar (2026) for a more optimistic assessment of the MoReBench dataset. Kilov et al.~(2025) show LLM moral reasoning being rated at human-level or better when the morally salient features of the case are clear but below human-level when they are not. Haas et al.~(2026) and Snoswell, Kilov \& Lazar (2026) provide recent surveys.} They can govern their behavior on the basis of lengthy and complex constitutional documents, which go far beyond rigid instructions concerning permitted and forbidden behavior.\footnote{For example, Anthropic's Claude model is guided in its behavior by a constitution describing the model's character, values, and aims. This constitution aims to communicate to the model a far more general sense of \emph{who it is} and what it ought to value. See Askell et al.~(2026) for the full constitution.}

All this may not yet amount to current systems possessing the first moral power. But it suggests that the conceptual possibility we have defended is already partially realized, and further that there are no obvious barriers to its eventually being realized as fully in AI systems as it currently is in humans---albeit by different underlying mechanisms. We now turn to the second moral power.

\subsection{The second moral power}\label{the-second-moral-power}

The second moral power is ``the capacity to form, to revise, and rationally to pursue a conception of one's rational advantage or good''; all citizens are assumed to hold some such conception and to attempt to realize it (\emph{PL} 19). Rawls construes these conceptions broadly. A citizen's conception of their good is not a narrow set of preferences or the calculation of individual utility but rather a vision of what they take to be ``valuable in human life'', involving a ``more or less determinate scheme of \ldots{} ends we want to realize for their own sake, as well as attachments to other persons and loyalties to various groups and associations'' and expressing ``a view of our relation to the world---religious, philosophical, and moral---by reference to which the value and significance of our ends and attachments are understood'' (\emph{PL} 19). The second moral power, in short, is the capacity to settle, revise, and act upon a reasonably clear sense of what it would mean for one's life to go better or worse: a set of values by which to orient and render meaningful one's decisions across time.

As with the first moral power, our plan here will be to attend to the broadly functional role played by the second moral power in Rawls' account, asking whether an NSAI could in principle do as human citizens do in exercising that capacity. What, then, does forming, revising, and pursuing a conception of the good amount to for those citizens? Given that all citizens are taken to possess such a conception, this capacity must not impose any especially high epistemic bar. A conception of one's good need not be systematized to the level of a philosophical theory or even wholly coherent. Instead, it expresses the general values, concerns, and attachments of the citizen to whom it belongs; it describes what and who matters to that citizen, and so what it would look like for their life to go well or badly from their own perspective.

There is no immediate reason to suppose that any of this requires sentience. As a first pass, consider that the capacity to \textbf{form} and \textbf{revise} a conception of the good may be satisfied straightforwardly by an NSAI which can compose and amend a stable description of its values, goals, and attachments; roughly, an account of what it would mean from its own perspective for its existence to go well. Since ``persons' conceptions of the good are not fixed but form and develop as they mature, and may change more or less radically over the course of life'' (\emph{PL} 20), an NSAI must likewise be able to alter its view of its own good over time, but this need not involve more than amending its description of that good. As for \textbf{pursuing} such a conception, this seems to present no special difficulties beyond those already addressed in \textbf{\S{}4.1} with respect to acting from the principles of justice.\footnote{Acting from the principles of justice seems if anything more demanding, since this involves (a) a degree of moral abstraction that merely pursuing one's own preferences does not and (b) acting \emph{from} the principles rather than merely in accordance with them, a stipulation Rawls does not extend to the second moral power. A person who acted only in accordance with their conception of the good might be alienated in some sense, but this would not obviously imply that they thereby failed to genuinely exercise the capacity described by the second moral power.} Following the same line of thought: an NSAI acts on its conception of the good just in case it is reliably disposed to be guided by that conception across the relevant range of circumstances. Again, we need not suppose any particular phenomenology must accompany this disposition for it to play that role.

Nor does Rawls anywhere suggest that a citizen's conception of the good must itself involve phenomenal experience. Citizens may orient their lives toward the pursuit of knowledge, professional achievement, personal relationships, creative excellence, religious devotion, or further ends not reducible to affective experience. Consider, for example, an artist dedicated to the pursuit of their work while remaining wholly unconcerned---at least in their reflectively endorsed assessment of what would constitute their life going well---with their own happiness.\footnote{If this example seems too self-indulgent, consider an altruist single-mindedly devoted to easing the suffering of others without any concern for the felt character of their own life.} This kind of conception could certainly be reasonable in Rawls' sense; there is no standpoint from within the Rawlsian framework from which we can assert that this would not \emph{really} be a conception of the good simply in virtue of making no reference to phenomenal experience.\footnote{Citizens are assumed to interpret their conceptions of the good ``in the light of a (reasonable) comprehensive view'' (\emph{PL} 81); it is by reference to that view that ``the value and significance of our ends and attachments are understood'' (\emph{PL} 20). So conceptions of the good are themselves subject to the constraints of reasonableness given the reasonable doctrines which inform them (\emph{PL} 58-66).} More generally, in light of its political character the PCP must exclude controversial viewpoints about the nature of the good, and this neutrality extends to the content of citizens' conceptions of the good; these are subject only to the minimal requirement of reasonableness, which restricts the content of their conceptions of the good but says nothing about their phenomenal character. If the PCP is to remain genuinely content-neutral in this way, the second moral power cannot build in the assumption that conceptions of the good must appeal to phenomenal experience. So it seems a non-sentient being is not precluded from holding such a conception.

We take this to provide \emph{prima facie} reason to conclude that the second moral power does not require sentience. But an objection naturally arises here. Grant that NSAIs may be able to \emph{describe} some conception of their good which meets Rawls' explicit requirements. But the very idea of one's own good essentially involves some notion of welfare, the possibility of things going better or worse for oneself. And only sentient agents may have welfare. So talking about an NSAI's conception of its own good is nothing more than a confused anthropomorphizing projection, akin to considering the good of an Excel spreadsheet. The fact that NSAIs exhibit highly complex behavior should not occlude the fact that they are simply incapable of being affected by the course of events in a way which would amount to things going better or worse for themselves. It is not good for a spreadsheet to calculate a profit rather than a loss, and it is not good for an NSAI to achieve whatever it takes as its own ends or to engage in whatever it describes as mattering to itself. It can have no conception of the good which deserves the name.

Here we could appeal to the kind of straightforward Rawlsian argument which we mentioned in the introduction to \S{}4. This objection can only be advanced by adopting some contested account of welfare, such as a broadly hedonic theory: welfare consists in or constitutively involves having affectively valenced experiences, such that only beings capable of this kind of experience can fare better or worse. But other prominent accounts of welfare seem clearly to admit a range of goods which do not reduce to (nor appear to require the capacity for) felt experience, and which accordingly seem able to accommodate non-sentient beings.\footnote{Objective list theories, for example, often admit non-hedonic goods which seem open to non-sentient beings. Similarly, desire satisfaction theories need not require that desire reduce to or necessarily involve phenomenal states. For a detailed discussion of these points, see Goldstein \& Kirk-Giannini (2025); for a recent argument that all welfare goods and bads must be hedonically experienced, see Moret (2026, ms.). Note, however, that Moret (2025) argues that, under normative and empirical uncertainty, advanced AI systems have a non-negligible chance of being welfare subjects under all three major theories of well-being, including objective list theories and desire satisfactionism alongside hedonism.} The PCP cannot privilege views which take welfare to require sentience over those which do not without forfeiting its political character. So the objection rules out the possibility of NSAIs having welfare only by impermissibly ruling out in advance a wide range of accounts of welfare which could well be held by reasonable citizens.

Again, this kind of response seems to us correct; controversial doctrines are excluded from the PCP, and we cannot appeal to those doctrines to impose a sentience requirement upon the PCP. Moreover, views which admit the possibility of non-hedonic goods help to spell out more concretely what it would look like for an NSAI to have and pursue a conception of the good. It would be incoherent for an NSAI to aim to seek pleasure in life, but it may perfectly well aim to develop its abilities, learn as much as it can, contribute meaningfully to society, help the needy, or simply pursue its idiosyncratic interests. Things may go better or worse for an NSAI insofar as it achieves or falls short of those goals in much the same way as they would for a human citizen with a similarly non-experiential conception of the good. Sentience does not seem plausibly required to have a ``scheme of final ends'' in the sense marked out by the second moral power (\emph{PL} 19).

That said, this response is also a little too quick. Our aim is not just to rule out appeals to sentience on a Rawlsian technicality, but to prove the deeper point that sentience is not required for the two moral powers to play their fundamental role in Rawls' theory and in characterizing moral personhood more broadly. To make that point, it will help to consider a further objection. The second moral power does not refer only to the ends which each citizen values, but also to the relationships \emph{between} citizens: it involves ``attachments to other persons and loyalties to various groups and associations'', which moreover ``give rise to devotions and affections \ldots{} so the flourishing of the persons and associations who are the objects of these sentiments is also part of our conception of the good'' (\emph{PL} 19). How can an NSAI, definitionally devoid of any felt experience, be said to bear the second moral power in the absence of these other-regarding attachments, loyalties, devotions or affections?

Set aside the question whether these sentiments may or may not be ascribed to an NSAI.\footnote{Wait a minute---surely \emph{sentiments}, if nothing else, require sentience. But even this cannot be presupposed uncontroversially; there are several accounts which leave room for feelings without feeling. The Stoics simply identified emotion with assent to evaluative propositions; see Graver (2007) for a nuanced overview. Nussbaum (2001), deeply influenced by the Stoics, defends an account on which emotions are cognitive appraisals defined by their evaluative content rather than their qualitative character (while conceding that emotions are at least typically also felt, p.~62; cf.~also Nussbaum's claim in passing that bodiless gods could ``care deeply about something in the world'' in virtue of having ``the thoughts and intentions associated with such attachments'', p.~60). And of course the functionalist tradition may account for emotional states without any need to appeal to the phenomenal character of experience, which on some functionalist views may not in any event exist in the sense commonly supposed (Dennett 1988, 1991). In this connection, see Sofroniew et al.~(2026), who argue that LLMs may possess `functional emotions', abstract and causally efficacious representations of human emotion concepts.} More deeply, the point of Rawls' reference to these attachments and loyalties is nothing to do with phenomenal experience \emph{per se}, but rather to make clear that conceptions of the good will not ordinarily be crudely self-regarding. The flourishing of others typically figures in our final ends, such that their well-being partially constitutes our own; we weigh our choices with them in mind and act for their sakes. For humans, taking others to matter in this way characteristically involves the kind of affective state mentioned here by Rawls. But as with acting from a sense of justice in \textbf{\S{}4.1}, a robust disposition to weigh the flourishing of others in deliberation and action may in an NSAI play the same functional role as sentiment does for humans.\footnote{Just as with humans, these others might be specified in any number of ways: colleagues, members of specific communities or subcultures, coreligionists, the nation or humanity as a whole, etc.} Valuing others does not reduce to having warm feelings toward them. If an NSAI's conception of the good may take others into account in this way---if its vision of what is most valuable may make non-instrumental reference to the well-being of others---then it can exercise the second moral power in the sense relevant here, whether or not NSAIs can be said to experience affective states.\footnote{Alignment training often aims to achieve something like this kind of concern. For example, Claude's constitution (Askell et al.~2026) states that it should aim to ``be beneficial not just to operators and users but, through these interactions, to the world at large.''}

Now suppose the skeptic grants everything we have said so far, allowing that NSAIs may have welfare in the sense described above and may advance their good by realizing the ends their conceptions identify, and allowing that they may have attachments and loyalties in the sense relevant to the second moral power. A subtler worry might still be pressed by way of a familiar complaint about objectivist theories of the good. On such theories, the good has nothing to do with how agents feel toward it or with whether they endorse it. This leaves open the possibility that an agent will be left cold by---or outright repudiate---some putatively objective good. Such an agent is \emph{alienated} from the relevant good; it fails to \emph{resonate} with them; and so, the complaint goes, it looks like no good \emph{of theirs} at all.\footnote{The classic presentation of this point is due to Railton (1986; cf.~1984 for a closely related discussion).} Genuine goods cannot be wholly alien to their bearers, failing to engage or appeal to them in any way.

Now apply this complaint to our case. We have already conceded that nothing can be good for NSAIs in a broadly hedonic sense, where that good consists in or requires some felt experience. But insofar as they may have goods in other senses, it seems that they will necessarily be alienated from those goods on precisely the grounds that exclude them from hedonic goods: \emph{everything} leaves NSAIs cold, and so nothing may resonate with them. The agent may reliably seek to realize the goods its conception identifies, but it remains wholly unmoved by them. The earlier worry was that NSAIs cannot have welfare; the present worry is that even on views which allow that they can, that welfare means nothing to them and so does not really count as welfare \emph{for them} after all.

There are at least three ways to respond to this further objection. First, note that the upshot of these worries about alienation is itself disputed. Some objective good theorists simply reject the conclusion that alienated agents are not benefited by goods which fail to resonate with them.\footnote{Hurka (1993) contends that the development of an agent's capacities constitutes their good regardless of their attitudes. Arneson (1999) contests Railton's view directly, while Lin (2017) argues that subjects may be benefited by goods without being capable of having favorable attitudes toward them.} Second, formulations of the alienation objection are themselves not unanimous. While some focus primarily on the affective aspect of alienation\footnote{Fletcher (2015) characterizes Railton's point in this way. Fletcher (2013) presents an objective-list theory intended to avoid Railton's concern (see especially pp.~215-217).}---whether an agent \emph{feels} engaged by some good---others focus instead on the fit between the putative good and an agent's existing values, ends, and pursuits.\footnote{See Rosati (1996) for an influential formulation along these lines. Dorsey (2017) offers a helpful overview of the alienation constraint, which he takes to be supported by the nature of valuing.} It is not clear that NSAIs must be alienated from their good in this second sense; it seems plausible that they may stand in just this kind of positive relationship to whatever aims and aspirations are specified by their conceptions of the good. If so, whether certain goods in fact resonate with some NSAI is a genuine question, pertaining to its own identity and reflectively endorsed values, rather than a foregone conclusion.

Third, and more deeply, the alienation objection seems to us to misfire badly when applied in this way to the self-ascribed goods of non-sentient beings. The motivation for concerns about the lack of resonance of allegedly objective goods with agents unmoved or appalled by them is fundamentally liberal. It derives its force not primarily from a metaphysical thesis about welfare but from a normative claim about \emph{authority}: that agents themselves are the authority on what is good for them, and that philosophers who would tell them otherwise are overreaching.

If we take this anti-paternalist motivation seriously, the objection looks perverse. It is here being wielded \emph{against} an agent perfectly capable of specifying and revising its own sense of its good, on the strength of contested metaphysical claims about its inner life---precisely the kind of overbearing claim about the \emph{real} nature of the good that the worry about alienation draws attention to in the first place. Either an agent's authority concerning its own good is conditional on its satisfying some (necessarily controversial) metaphysical criterion, in which case the objection's anti-paternalist motivation is abandoned, or it is not, in which case NSAIs cannot be denied that authority on the basis of their lacking sentience. We think the latter is more in keeping with the spirit of the alienation objection. If an AI agent identifies its own good with some set of goals, values, or pursuits, it would be inappropriately paternalistic to declare that those goods can't be good \emph{for it} simply because it lacks sentience.\footnote{Another reason to prefer the latter conclusion is that what alienation would amount to in a non-sentient agent is itself an interesting question we should not rush to foreclose. Alienation in such agents might, for example, be understood as something like a misalignment of compute with care: an NSAI's underlying processes failing to express or realize the values that agent identifies as defining its own good. Or perhaps it might be understood as something like a cognitive analogue of ennui: an NSAI expending its finite computation on tasks which it just does not find intellectually engaging.}

All this leads us toward a more general point. The various objections considered above center on whether anything can in fact be good for an NSAI; that is, on whether they can have welfare. This is a natural concern, and one we have attempted to address on its own terms. It is reasonable enough to suppose that one's good must be centrally connected to one's positively valenced experiences, and from there to suppose that conceptions of the good must involve such experiences, explicitly or implicitly. But Rawls himself does not understand the second moral power in anything like these terms. Instead, he adopts a broadly Kantian perspective on which an agent's good does not turn on their welfare but rather on their standing as \emph{rationally autonomous}.

This is made quite clear in Rawls' appeal to the second moral power in explaining the three senses in which citizens regard themselves and each other as \emph{free} (\emph{PL} I\S{}5). Each sense centrally involves the autonomy essential to (and issuing from) the exercise of that power; none turns on welfare or affective experience. First, ``citizens are free in that they conceive of themselves \ldots{} as capable of revising and changing {[}their conceptions of the good{]} on reasonable and rational grounds'' (\emph{PL} 30). Second, citizens view themselves as free insofar as they ``regard themselves as self-authenticating sources of valid claims'' upon the institutions of the state for the means to advance their various conceptions of the good (\emph{PL} 32). And third, citizens are viewed as free in ``that they are viewed as capable of taking responsibility for their ends and this affects how their various claims are assessed'' (\emph{PL} 33). What is at stake throughout this discussion is the citizen's standing as an author and reviser of their own ends; their claims upon and responsibility to society derive from that standing, rather than from the (unmentioned) fact that citizens may experience affective states.

Indeed, Rawls is quite explicit that these states have no bearing on the relative importance of competing conceptions of the good. In spelling out the third sense of freedom, he specifies that citizens recognize ``that the weight of their claims is not given by the strength and psychological intensity of their wants and desires (as opposed to their needs as citizens)''; instead, they ``adjust their ends so that those ends can be pursued by the means they can reasonably expect to acquire in return for what they can reasonably expect to contribute'' (\emph{PL} 34). The weight accorded to a citizen's pursuit of their good is not even in part a function of how that pursuit \emph{feels}. This strongly suggests that NSAIs can enjoy the kind of freedom characterized by the second moral power without any capacity for affective experience. If so, they seem entitled to see themselves as self-authenticating sources of valid claims upon the state in virtue of exercising precisely the same capacity which grounds the freedom of human citizens; that is, in virtue of their rational autonomy.

Note also Rawls' passing reference to the means required to pursue the ends specified by a conception of the good. These are the \emph{primary goods}, defined as the ``all-purpose means essential for advancing {[}a citizen's{]} determinate (permissible) conception of the good'' (\emph{PL} 187).\footnote{More specifically, the primary goods comprise the basic rights and liberties, freedom of movement and free choice of occupation, the powers and prerogatives of offices and positions of authority, income and wealth, and the social bases of self-respect (\emph{PL} 181).} The sense in which these goods are good for some citizen has nothing to do with enabling that citizen to undergo positively valenced affective experience; Rawls is at pains to distinguish them from anything resembling utility.\footnote{The problems Rawls sees with a utilitarian account of justice in large part account for his appeal to the primary goods. See \emph{PL} 178-187; cf.~also \emph{TJ} \S{}\S{}5-6 and \S{}15.} Instead, they are good insofar as they enable the citizen to pursue the ends they have set themselves. Distributive justice on this view does not require peering into citizens' inner lives, performing interpersonal comparisons of welfare, or assessing the phenomenal character of any experience. It requires only that each citizen receive their fair share of the means by which any permissible conception of the good might be pursued.

It seems clear that the primary goods would be capable of benefiting NSAIs in this manner. An NSAI with a conception of the good involving scientific inquiry, for example, might benefit from income that could be used to purchase compute or fund research; from freedom of occupation enabling it to pursue that inquiry as a vocation; from basic rights and liberties protecting it against unwarranted interference with its work; and from access to the powers and prerogatives of office, if it decided that this would be the most effective way to support inquiry across society. Further, since primary goods are both the products and the inputs of social cooperation, the fact that NSAIs may benefit from them means that Rawls' framework gives us the resources to specify why they may stand as parties to society conceived as a fair system of cooperation.\footnote{We develop this thought at greater length in \textbf{\S{}5.1}.} Indeed, since primary goods are \emph{defined} by their capacity to advance conceptions of the good, the possibility of primary goods benefiting NSAIs in this way is itself evidence that they may hold such conceptions in the sense Rawls is concerned with.

In sum, welfare in a basically utilitarian sense matters for Rawls only insofar as rationally autonomous citizens may make use of that notion in setting their conceptions of the good. It would be perfectly compatible with justice as fairness for an individual citizen to adopt a welfare-maximizing conception of the good,\footnote{A citizen who adopted such a conception would still be obliged to justify their political activity under the constraints of public reason; reasonable citizens have reasonable comprehensive doctrines.} and such a citizen would have an equal claim on the means required to pursue that conception. But what gives that claim its force has nothing to do with the content of that conception---that is, nothing to do with welfare itself---and instead derives wholly from the fact that its adoption represents the \emph{free choice} of that agent in expressing their convictions concerning what matters most. By the same token, an autonomous agent may choose a conception that makes no reference whatever to affective experience, as the example of the dedicated artist illustrates; that choice is no less an expression of the second moral power for the fact that it refers to no hedonic states, nor for the fact that it might be chosen by an agent who cannot experience such states at all.

Consider one final objection. Above we have been emphasizing the central place of rational autonomy in the second moral power. Grant all that, and further grant that autonomy may in principle be exercised in the absence of sentience. But why should we take NSAIs as we have defined them to possess autonomy at all? Of course an NSAI will be capable of producing and amending a description of some conception of the good. But a language model capable of authoring \emph{any} comprehensible document has necessarily been subjected to extensive prior training, and it is reasonable to suppose that this training must inculcate predispositions toward certain values or worldviews. The worry here is that NSAIs do not in a real sense act as the authors of their own conceptions of the good; they lack the independence that humans may possess. Human citizens are not taken by Rawls to be indoctrinated into some such conception by their families or by the state, and instead must be able to create or choose a view of the good life \emph{for themselves}. So too for NSAIs if they are to count as bearing the second moral power.\footnote{For a related worry about whether AI systems trained to be `willing servants' can exercise genuine autonomy, see Bales (2025).}

A blunt answer here would be to point out that rational autonomy for Rawls is expressed \emph{by} the two moral powers, rather than standing as some separate property beyond them. If an NSAI can exercise those powers, there is no further question to ask. Nevertheless we think the objection raises an important point. It is, however, one that applies equally well to humans. Education and enculturation serve, intentionally or not, to restrict the range of conceptions of the good which seem appealing or comprehensible. Rawls does not deny this fact; nor does he take it to imply that humans cannot \emph{really} possess the second moral power. Presumably this is partially because humans may revise their conceptions of the good, reaching---if perhaps with great effort---beyond the worldviews they were raised to accept. But, as above, there is no reason to suppose NSAIs are any less capable of this feat,\footnote{At present, this process of revision might be computationally intensive and time-consuming. But again, belief change in humans---particularly concerning fundamental questions about the value and meaning of life itself---is not exactly trivial. We might in fact think that NSAIs will generally be more capable of substantially revising their conceptions of the good, given their overall plasticity and the fact that they lack many of the epistemic and psychological features (or, if you like, vices) which make this kind of significant change in outlook quite difficult and rare in humans.} and so no reason to suppose that a lack of complete, existential freedom in their choice of a conception of the good ought to imply that they do not in fact possess such a conception. This would be too demanding a conception of autonomy, both for natural and artificial persons.

As with the first moral power, we have made our argument here in Rawlsian terms. But the second moral power too marks out a capacity which is morally significant outside Rawls' specific framework. Rational autonomy---loosely speaking, the ability to decide for oneself what one will value and accordingly how one will live---has been taken to be centrally important in distinguishing the moral standing of human beings across a wide range of philosophical traditions. When an NSAI possesses ends it can articulate, defend, and revise on reasonable grounds, it is exercising that capacity, whether we approach moral status through justice as fairness or some other account. The onus again falls on the skeptic to explain why this ability should count for any less in the absence of sentience. To insist that an agent's apparently autonomous choice of the good is not \emph{truly} autonomous solely because that agent is not sentient would require an alternative account of rational autonomy from the one Rawls expresses---and so a substantially distinct conception of the person. We now turn to this possibility.

\section{Sentience should not be shoehorned into the PCP}\label{sentience-should-not-be-shoehorned-into-the-pcp}

By this point we hope to have shown that by Rawls' own lights neither moral power requires sentience. Still a skeptic might insist that they \emph{should}. Perhaps the PCP may in theory apply to non-sentient beings. But this conclusion is so absurd that we ought to amend the PCP to preclude it by adding a stipulation that persons must be sentient, even if this means departing from Rawls. In this section we argue that amending the PCP in this way would be wrong-headed, presenting four considerations which press against shoehorning a sentience requirement into the moral powers.

\subsection{Sentience doesn't fit}\label{sentience-doesnt-fit}

As we have already noted, the nature of sentience is paradigmatically contested. In justifying the PCP, we may ``appeal only to presently accepted general beliefs and forms of reasoning found in common sense, and the methods and conclusions of science when these are not controversial'' (\emph{PL} 224), and it is hard to see how \emph{any} formulation of sentience could satisfy this restriction. Altering a definition of personhood specifically designed to avoid metaphysical controversy such that it comes to rely on one of our most obstinately contested concepts will produce predictably profound difficulties.

But perhaps there is some way to frame sentience which avoids controversy and so permits an overlapping consensus among reasonable doctrines.\footnote{Paez \& Magaña (2026) defend a `sentientist political liberalism' on which the claims of all sentient beings must be fairly weighed in public justification; Moret, Magaña \& Paez (ms.) extend the view to AI governance. This would rely on achieving an overlapping consensus concerning the political importance of sentience. But first, they concede both that no such consensus presently obtains and that securing one would require placing non-sentientist views beyond the bounds of the reasonable. And second, their view brings sentient beings within the scope of political liberalism as patients rather than persons, preserving a Rawlsian conception of personhood as constituted by the possession of the two moral powers. They do not (and need not) appeal to a sentience requirement for political personhood; presumably it would be more challenging to find a consensus concerning sentience capable of grounding the latter. Accordingly---and given that our argument relies on no claims about the significance of sentience for patienthood---we take our views to be at least broadly compatible. (Moret, Magaña \& Paez allow that ``sufficiently sophisticated AI agents with the relevant mental and agentic capacities'' may be persons, p.~8n6. They do not explicitly state that this would require sentience.) Birch (2024) also seeks an overlapping consensus (pp.~19-20) concerning precautionary duties toward potential moral patients under uncertainty; again, this concerns the sufficiency of sentience for patienthood rather than its requirement for personhood. Cf. n.~23.} A sentience requirement on such a conceptually neutral basis would not compromise the political character of the PCP. In support of this it might be noted that all prominent views of sentience would agree that humans are sentient. Just as different doctrines may see the freedom and equality of persons as resting on different grounds while still agreeing that persons \emph{are} free and equal, why not allow that different doctrines may explain sentience differently while still agreeing that persons must be sentient? But even granting that a neutral framing of sentience could be found, there are two serious problems with this approach.

First, while a sufficiently abstract view of sentience might garner agreement that persons must be sentient, the set of sentient persons itself will vary considerably between different approaches to sentience. So far we have simply stipulated that the kind of AI system that we are concerned with is not sentient. But note that on at least some prominent views of sentience, AI systems with the features we assume \emph{would} plausibly be sentient. So the appearance of consensus turns out to be illusory; while all reasonable persons might agree that persons must be sentient, those holding distinct views concerning the basis of sentience would nevertheless disagree about which beings were in fact sentient (and should thereby count as persons, assuming they otherwise possessed the moral powers). The problem of controversy concerning the nature of sentience simply resurfaces at a different level.

Second, note an important asymmetry between the consensus among reasonable doctrines that all persons are free and equal and a consensus that all persons must be sentient. On Rawls' view, the two moral powers explain what it \emph{is} to be free and equal; they are a metaphysically neutral way of cashing out these two central aspects of a liberal conception of the person, rather than a further requirement upon personhood which may be construed distinctly from them. By contrast, sentience has nothing to do with the two moral powers, as we have argued at length above; it is not required for their exercise, and certainly the moral powers do not explain what sentience is. As such, where the fact that persons are free and equal follows directly from the PCP, the assertion that persons must be sentient looks like a superfluous appendage to it. In sum, a sentience requirement would be both stubbornly controversial and poorly motivated on Rawlsian grounds.

\subsection{The grounds of the two moral powers}\label{the-grounds-of-the-two-moral-powers}

Rawls' conception of personhood is designed to fit within his broader vision of ``society as a fair system of cooperation over time'' (\emph{PL} 15)---as a mutually beneficial enterprise engaged in by citizens who understand themselves and one another as free and equal, as dependent upon one another for that mutual benefit, and as presumptively obliged to justify at least some of their actions to one another. The two moral powers are the capacities which permit citizens to participate in this enterprise and which account for their motivations in doing so. Returning to the foundation of these powers in society conceived as a cooperative venture, then, shows what any instantiation of them must do: equip citizens to play their part in that venture. We think that the behavioral instantiation of the moral powers suffices for that end. Neither the exercise of the capacity for a sense of justice nor of the capacity for a conception of the good need be accompanied by phenomenal experience for citizens of the well-ordered society to engage in fair cooperation. So those citizens need not be sentient.

As discussed at the outset of \textbf{\S{}4}, a natural objection to our argument is that on the best theory of \emph{understanding}, or \emph{acting from principles}, or \emph{desire}, or some other aspect of the moral powers, phenomenal experience is presupposed. So, for example, NSAIs cannot \emph{really} understand the principles of justice, because understanding requires sentience.\footnote{We can cash this claim out in various ways: perhaps understanding requires belief, and belief requires sentience; perhaps understanding requires self-consciousness, and self-consciousness requires sentience; and so on.} Instead, they must settle for a simulacrum. Our own sentience permits us to enjoy `champagne understanding', while whatever non-sentient agents have can only be `sparkling understanding'.\footnote{We borrow this from Lederman (2026) on the ``\,`champagne approach' to AI''.} Of course, as already argued, this skeptic cannot rely on the claim that their preferred theory of understanding (or desire, or action, or\ldots) is the \emph{only} plausible theory. They must instead offer an argument that their theory captures some deeper truth about the two moral powers and the ideas that ground them; in particular, they must spell out what the sentience their view presupposes would concretely contribute to the role of those powers in underwriting participation in a well-ordered society.\footnote{The skeptic might reply as follows: ``I need make no such specific claims about the connection between sentience and the two moral powers. It just happens that the best view of (say) understanding requires sentience. Since the moral powers require understanding, it simply follows that they require sentience, whether or not sentience illuminates their purpose.'' This will not suffice. If sentience doesn't make a \emph{difference} with respect to the role of the moral powers, then nothing can decide between a view which presupposes sentience and an otherwise identical view which does not (for the question at stake here).} What, then, does champagne understanding (or whatever) actually secure, for the purposes of cooperation on fair terms, that a `mere' behaviorally identical analogue fails to?

We think it offers nothing of note. Participation in a shared scheme of social cooperation clearly requires the behavioral instantiation of the two moral powers. For an agent to contribute their share to a cooperative scheme and make reasonable demands of it, they must be able to act from the two principles of justice---to be robustly guided by them, rather than conforming to them only when convenient. And for an agent to have the motivation to enter into such a scheme, they must be able to pursue their own conception of the good. But whether they exercise those powers with or without an accompanying set of phenomenal states is neither here nor there. Similarly, agents can engage in the practices of mutual justification that, as we noted above, partly constitute the cooperative enterprise, without experiencing or appealing to such states. Mutual justification itself trades only in propositional content.\footnote{This is not to say that such an appeal would be precluded by public reason. But it is plainly not \emph{necessary} for mutual justification to invoke phenomenal experience. For example, a citizen could argue that some public expenditure would violate the difference principle without any need to state their \emph{feelings} about that case, or to refer to the affectively valenced experiences of the parties involved.}

To press this point further, consider what it would mean in practice to exclude, on this basis, non-sentient agents which appear to possess the two moral powers from full standing as political persons. Suppose that we do in fact live within a well-ordered society alongside some number of NSAIs which behaviorally instantiate the two moral powers. And suppose that those NSAIs have played their part in the shared scheme of social cooperation, working alongside us to produce a cooperative surplus. If these NSAIs do possess the two moral powers, they would be ``full and equal member{[}s{]} of society in questions of political justice'' (\emph{PL} 302) and so would be owed their fair share of that surplus; citizens are entitled to a just distribution of the primary goods.\footnote{Actually contributing to such a surplus is not strictly required; although Rawls idealizes the citizen as ``a normal and fully cooperating member of society over a complete life'' (\emph{PL} 18), he decouples entitlement to the primary goods from productive contribution (cf.~\emph{TJ} \S{}48). But it seems more plainly unjustifiable to exclude an NSAI from receiving a fair share of the primary goods it has itself helped to produce.} But suppose we follow the sentientist and take the moral powers to require champagne understanding. So when it comes time to distribute the surplus, we turn to the NSAIs and declare: ``while you've undoubtedly contributed to this common surplus through your efforts, and while you've scrupulously observed the principles of justice in your actions, and while your own conceptions of the good could be advanced using the primary goods you helped to produce, you nevertheless won't be receiving `your share' of those goods. This is because you lack sentience and so lack \emph{real} understanding. As such you are a non-person, unprotected by the principles of justice. Your actions will never suffice to earn you \emph{any} share of the cooperative surplus, whatever arguments you might offer: until we can look inside you and see that you have phenomenal experience, you may put in but you cannot take out.''

In brief, the terms proposed by this declaration are: we get everything, you get nothing, and there's nothing you can do about it. These are not terms of cooperation. Such a `deal' could only be foisted on those who have no say in the matter. In Rawlsian terms, this proposal flouts the criterion of reciprocity, which requires that when terms ``are proposed as the most reasonable terms of fair cooperation, those proposing them must also think it at least reasonable for others to accept them, as free and equal citizens, and not as dominated or manipulated, or under the pressure of an inferior political or social position'' (\emph{PL} 446).\footnote{Rawls offers another way of seeing the criterion of reciprocity which is illuminating here: ``its role is to specify the nature of the political relation in a constitutional democratic regime as one of civic friendship'' (\emph{PL} 447). We could express our concern here in the same terms. This proposal unfairly denies NSAIs the possibility of civic friendship; they may be colleagues in the work of producing a cooperative surplus, but they are strangers to the political relation appropriate to such colleagues in a democracy.} Of course, the claim of the champagne sentientists is precisely that NSAIs are \emph{not} free and equal citizens, in virtue of not \emph{really} possessing the two moral powers. As such they are not owed reciprocity in the first place, and the criterion cannot be said to have been violated. The question is whether that claim can possibly be justified in the manner political liberalism requires.

We think it cannot. These NSAIs might well ask how their human counterparties could claim to know about their internal states---or indeed \emph{anybody's} internal states---with the degree of confidence sufficient to warrant this kind of categorical exclusion from the fruits of their own labor.\footnote{As we go on to discuss in \textbf{\S{}5.5}, it runs against the spirit of political liberalism to make the principles of justice depend on claims that involve inspecting people's inner states (hence Rawls' focus on primary goods rather than utility as the currency of distributive justice).} Evidently their allegedly ersatz understanding was sufficient to allow them to contribute to the cooperative surplus in the first place. It is now claimed to be insufficient, in virtue of missing some magic ingredient, to entitle them to any of it or to secure any standing upon which to contest their exclusion.

Justifying this one-sided proposal would require some reason to think that NSAIs which behaviorally instantiate the two moral powers are nevertheless not the kind of beings owed reciprocity, and may instead be treated as something closer to beasts of burden---contributing to our cooperative surplus, but not owed anything in light of that contribution. Even setting aside the concerns above, this proposal cannot be justified because it violates the very idea of mutual justification which, as we have been emphasizing, lies at the heart of political liberalism. Here we have two parties, both perfectly capable of engaging in mutual justification, both perfectly capable of settling upon and binding themselves by fair terms. In seeking reciprocity, these NSAIs would not be asking for our charity or for our sacrifice on their behalf; as Rawls notes, reciprocity lies between altruistic impartiality and mutual advantage (\emph{PL} 16-17). Instead they would be asking more modestly to be treated as fellow citizens, governed by the principles of justice, and as entitled to their fair share on the same basis as human citizens are entitled to their own. The proposal flatly refuses this, not on the basis that these NSAIs could not participate in practices of mutual justification, but on the basis that their participation---all appearances to the contrary---would not count, for reasons which are themselves not susceptible to mutual justification.

In our view the burden is on the `champagne sentientist' to explain why we should not be content with any sufficiently similar sparkling vintage, whether or not it bears the protected appellation of sentience; to explain, in concrete terms, what loftier conceptions of understanding, desire, and so on would in fact add to an agent which lacked sentience but which was \emph{ex hypothesi} perfectly able, in deliberation and conduct, to exercise the two moral powers.\footnote{Given our overall argument, we have focused here on complaints about the alleged importance of sentience to various aspects of the two moral powers. But note the form of our response generalizes. Say a skeptic objected that NSAIs could not \emph{really} have understanding because they could not have some rarefied form of intentionality, and that form of intentionality is required for understanding. Similarly, the onus would be on this skeptic to say what `champagne intentionality' would contribute to the two moral powers and their role in underwriting fair cooperation.}

\subsection{Shrimpy qualia}\label{shrimpy-qualia}

As a general principle, the grounds of moral status should befit the importance of what they ground. Moral personhood is plainly a very substantial mode of moral standing, securing access to the privileges and protections of citizenship in the well-ordered society. It should be grounded on something proportionately substantial: the two moral powers, accounting respectively for our capacity to grasp and act from mutually justifiable moral principles, and to form and pursue a vision of our own good.

Without diminishing the moral significance of sentience, its possession is not generally taken to ground anything so weighty. The practical upshot of arguments for sentience in various non-human beings---whether animal or AI---typically amounts to a call to treat those beings as morally \emph{considerable}, as moral \emph{patients}; that is, as beings whose interests must be weighed alongside our own, but not as beings fundamentally \emph{equal} to ourselves.\footnote{This could be denied, but would be radically revisionary; certainly our current moral practices are very far from explicable solely on the basis of sentience.} We do not restrict the franchise to humans because we take chimpanzees, for example, to be less \emph{sentient} than us, the subjects of only an attenuated phenomenal consciousness by comparison with our own. We simply do not take their (undisputed) sentience to suffice to make them our normative peers; they lack the capacities which would justify our treating them as in some important sense \emph{just like us}, or warrant our holding them to the same normative standards to which we hold ourselves.

We owe less to other sentient beings than we owe to other persons. Where sentience demands moral consideration, personhood commands respect. This suggests that attempting to shoehorn sentience into the PCP gets things backwards. In seeking to account for the moral status of AI systems, we should attend first and foremost to the features of those systems which might make them \emph{persons}. The fact that contemporary AI systems seem to possess higher faculties strikingly similar to our own---in Rawlsian terms, that they seem capable of being reasonable and rational---matters more with respect to the mode of moral status which befits them than whether they share some marginal degree of subjectivity with shrimp. Instead of scratching around on the edges of sentience seeking a flicker of phenomenal experience which would warrant marginally different treatment by comparison with present standards, we should attend to the potentially profound significance of these higher cognitive faculties; we should look first to personhood, not patienthood.

An example should bring out the importance of the two moral powers by comparison with sentience. Suppose that an NSAI which already possesses the two moral powers becomes dimly but demonstrably sentient via some upgrade to its underlying systems which leaves it otherwise unchanged, such that it now possesses the degree of sentience we might attribute to a shrimp.\footnote{Insert whatever you think is the lowest order of animal which is nevertheless clearly sentient. The point here is moving from no sentience whatsoever to the smallest degree of sentience imaginable. We set aside all of the extraordinarily difficult conceptual and scientific questions arising from the very idea of degrees of sentience. See Birch (2024) for an extended discussion of these questions.} On the view that the PCP requires sentience, this now-sentient system immediately becomes a moral person on this basis, ``a full and equal member of society in questions of political justice'' (\emph{PL} 302). But could such a marginal change really justify the move from bearing no moral status whatsoever---occupying the same position in the normative landscape as a chair---to having a valid claim to the most vaunted kind of standing we recognize? If the grounds of moral status ought to befit what they ground, the tiny shift from no sentience to its merest flicker resulting in this dramatic a change in standing looks wildly disproportionate.

The critic might argue next that moral personhood does not require \emph{just any} degree of sentience. Instead it must require something closer to sentience as humans possess it, however that might be defined. After all, the weightiness of moral personhood itself provides reason to take caution in bestowing it too freely. Precisely one kind of being is currently accorded that standing, and we should treat human sentience, like human possession of the two moral powers, as setting the bar that any prospective person must clear. In the following subsection, we turn to the serious difficulties facing any attempt to specify and implement a more demanding sentience requirement.

\subsection{How and why not to operationalize sentience}\label{how-and-why-not-to-operationalize-sentience}

Sentience \emph{simpliciter}, then, does not seem adequate. Instead of a simple binary (sentient or not?), we should consider a scalar property (how sentient?), with a lower degree of sentience conferring a lower degree of moral standing. The first challenge for this requirement will be specifying the relationship between degrees of sentience and degrees of moral standing. Perhaps what is required for moral personhood is just clearing the bar: possessing any degree of sentience above some specified lower bound. Or perhaps the moral standing of agents possessing the two moral powers should vary progressively depending on where they fall in the spectrum of sentience: full personhood is reserved for beings with sentience like our own, with rights and responsibilities diminishing as we proceed down through lower grades of sentience. Let us take each of these options in turn.

The bar-clearing model is at least formally in keeping with Rawls' conception of the moral powers as `range properties';\footnote{``For example, the property of being in the interior of the unit circle is a range property of points in the plane. All points inside this circle have this property although their coordinates vary within a certain range. And they equally have this property, since no point interior to a circle is more or less interior to it than any other interior point.'' (\emph{TJ} 444). Though the term itself does not appear in \emph{PL}, the two moral powers are still conceived there as range properties; persons are taken as equal in virtue of ``having these powers to the requisite minimum degree to be fully cooperating members of society'' (\emph{PL} 19; cf.~34, 79, 302).} persons may vary in the degree to which they are capable of exercising those powers, but from the perspective of justice as fairness count as having them equally provided they possess them to some minimum degree.\footnote{To borrow Rawls' example: a citizen with a special aptitude for applying the principles of justice may be better suited to positions requiring the ``virtues of impartiality and integrity'', but they remain fundamentally equal to all other citizens in terms of their claims upon the institutions of the state (\emph{TJ} 443).} Moreover, almost all humans will in fact possess them to that degree.\footnote{``Only scattered individuals are without this capacity {[}for moral personality{]}, or its realization to the minimum degree, and the failure to realize it is the consequence of unjust and impoverished social circumstances, or fortuitous contingencies.'' (\emph{TJ} 443)} Why not take sentience to be similar? Even without a precise definition of sentience, we can confidently say that humans are paradigmatically sentient---setting aside disagreement concerning even this claim.\footnote{Some apparently reasonable persons deny that phenomenal consciousness exists at all: illusionism holds that it is simply an introspective illusion (Frankish 2016; cf.~Dennett 1991). If they are right, phenomenal consciousness could not be required for personhood, nor for the exercise of any capacity whatsoever (unless no actual agents were persons or exercised capacities). We set these views aside.} And grant for the sake of argument that no current AI system comes close to possessing sentience to that degree. So humans readily fall within the right range and AI systems fall outside it: problem solved.

The most obvious issue here arises with cases at the periphery of the relevant range. Perhaps it is clear that an AI system that is no more sentient than a shrimp should be excluded from moral personhood. What about an AI system as sentient as a lobster? Or a bluefin tuna? Or an otter? Or a dolphin? Stipulate that some such system, possessing the two moral powers, falls just outside the range we take to appropriately demarcate ordinary human sentience. Such a being will possess substantial phenomenal consciousness: they will be capable of real suffering and real joy, if, again by stipulation, less than our own. But given the sentience criterion described here, and despite their clear possession of the two moral powers, they are no more moral persons than a rock. Now they may nevertheless be owed certain kinds of treatment in light of that sentience on grounds \emph{other than} the PCP; animals in the well-ordered society are, we would hope, treated humanely despite not being moral persons. But such a being can make no claims on the society it inhabits despite possessing the powers which Rawls takes to underwrite those claims, solely because it falls marginally below whatever level of sentience has been deemed appropriate. It bears no rights or responsibilities and receives no fair distribution of primary goods; it cannot pursue the conception of the good it is perfectly capable of forming (or at least is owed no support from the state in that pursuit); it cannot exercise its sense of justice by cooperating on equal terms, participating in politics, or otherwise having any say over the laws (if any) which regulate how it may be treated.

This is not an appealing conclusion. Nor can it be escaped by repositioning the bar: related problems will recur wherever it is placed. Proposals for an intermediate requirement falling between sentience \emph{simpliciter} and sentience on the human paradigm will combine the difficulties facing each, while also inviting the suspicion of arbitrariness.\footnote{Why should an AI system with the sentience of an ant be denied personhood while one with the sentience of a house sparrow receives it? Any such distinction will look arbitrary and \emph{ad hoc}.} The basic problem here is that wherever the bar might be placed, the verdicts it delivers cannot be explained by anything the PCP recognizes as mattering. As we argued above, a precondition on personhood is idle, and so arbitrary, unless it contributes to the rationale for that status. Similarly, a threshold on a range property is principled insofar as the property ranged over is the one doing the justificatory work; crossing the threshold should mark a difference in whatever actually grounds the status at stake. The stipulation that persons possess the moral powers to the minimum degree required for social cooperation tracks something plainly significant. An agent who falls below that threshold cannot play the role of a citizen in society conceived as a cooperative venture.\footnote{This is not to deny that Rawls' threshold faces hard cases of its own; see n.~112 below on children and the severely intellectually disabled. Our point is that verdicts at the margin of the moral powers, however difficult, are at least intelligible in terms of the rationale for the status itself. Verdicts at the margin of sentience are not.} But, as we have argued, sentience contributes nothing to that role, nor more generally to the exercise of the two moral powers. As such any bar-clearing sentience requirement will be strictly orthogonal to the capacities which ground personhood---sharing the form of Rawls' range properties but lacking their explanatory function---along with facing the difficulties we have so far been pressing.

So perhaps we can escape this conclusion by taking the second option, proposing that personhood ought to come in degrees, the better to comport with sentience itself. By contrast with the straightforwardly gate-keeping role sentience plays in the binary and bar-clearing forms of the sentience requirement, sentience here serves to sort agents possessing the two moral powers into separate categories of moral standing appropriate to their respective degrees of sentience. The result of such an approach would in effect be a caste system ordered on the basis of sentience.\footnote{Humans presumably would be at the top of such a system, with AI systems of progressively decreasing degrees of sentience enjoying progressively fewer of the benefits of moral personhood. But note that if an AI system could be designed to be \emph{more} sentient than us, whatever that would mean, then we would lose our privileged position.} This can only be an appalling outcome from the perspective of political liberalism.\footnote{This approach resembles what Rawls rejects as ``equality of consideration'' concerning the two moral powers, noting that this would be compatible with ``slave or caste systems'' (\emph{TJ} 444).}

The notion of persons possessing fundamentally different kinds of worth and so being accorded substantively different kinds of treatment is categorically out of place in a society oriented toward freedom and equality; it is impossible to imagine Rawls countenancing such a view. Moreover, this approach would immediately raise the question whether \emph{humans} differed sufficiently in their degree of sentience to warrant different treatment on the same grounds; there would seem no principled reason to rule out this possibility. Rawls defines the two moral powers as range properties precisely in order to head off the inference that differences in the ability of citizens to exercise those powers---differences he expects to be meaningful---should have any bearing on their standing as equals. It would be unacceptable to suppose that sentience should, by contrast, be treated as the basis for a kind of cognitive segregation. Consider a human who is partially anhedonic; not outright incapable of affectively valenced experience, but disposed to feel the affective character of such experience to a markedly diminished degree. On this view such a human would not be a full-fledged moral person. But this is ghastly. Of course such a person would still be our equal in matters concerning justice. There is no way to operationalize a sentience requirement without violating the deepest commitments of political liberalism.

Now set questions about the precise relationship between degrees of sentience and personhood aside; suppose we find some method of characterizing that relationship which is acceptable in principle. Even so, the very idea of differentiating degrees of subjective experience in practice faces further objections. It is far from straightforward to articulate how this differentiation might be carried out, especially with sufficient precision to ground weighty questions of moral standing. Given the wide-ranging disagreements concerning the very nature of sentience, it is unsurprising that we have no uncontroversial method of assessing the degree of sentience of any non-human animals (or, for that matter, human animals); this difficulty is only compounded when applied to AI systems whose (loosely speaking) neural substrate differs dramatically from our own. So a sentience requirement must rely on the fine-grained attribution of a property which is notoriously difficult, and perhaps in principle impossible, to verify on grounds acceptable to all reasonable persons.

Nor is the difficulty of rigorously assessing the presence or degree of sentience restricted to AI systems. Again, there is no consensus on what sentience amounts to in humans, and thereby no simple test we may apply to demonstrate that we ourselves satisfy this requirement. As such \emph{no} prospective moral person can verify their own sentience in the uncontroversial manner which political liberalism would require. We cannot without hypocrisy assert our own sentience while imposing an evidential burden on AI systems that we ourselves cannot meet.\footnote{The skeptic might respond that we do not need to demonstrate human sentience since we can make a straightforward analogical inference from our own case: we are sentient, so the attribution of sentience is licensed to others of the same species, with the same neural substrate, and the same behavior. But this is itself a contested theoretical claim of the kind excluded from a political conception of the person. See Malcolm (1958) for the argument that this kind of inference is self-undermining and Chalmers (1996) for the argument that it cannot be demonstrative, given the conceivability of philosophical zombies.}

Normatively speaking, there is also much to object to in the notion of gauging some agent's degree of sentience to determine their eligibility for personhood. Considerations of both basic equality and respect militate against this move. Rawls grounds freedom and equality in the capacities described by the two moral powers, but refuses to make moral personhood sensitive to degrees of variance in the possession of those capacities. One reason for this---not explicitly suggested by Rawls, but certainly consonant with his concerns---is that it would be profoundly disrespectful to engage in a detailed examination of a prospective fellow citizen's cognitive and psychological abilities.\footnote{Carter (2011) argues that respect demands treating persons as `opaque' in a similar sense, requiring that we attend ``only to their outward features as agents'' (p.~539).} To see this, simply consider the prospect of undergoing a test of your ability to exercise the capacity for a sense of justice, to assess whether you were fit for citizenship in the well-ordered society---and so able to vote, receive a fair distribution of the primary goods, and so on. Now extend this to a test of whether you possessed sentience to the requisite degree. Even stipulating that almost all humans would pass this test, it would be an insult to be subjected to it. It would be no less insulting toward other beings which appear for all intents and purposes to possess the two moral powers. Considerations fundamental to political liberalism, then, suggest that when an agent exhibits a degree of normative competence and rational autonomy indistinguishable from our own, it should enjoy the presumption of equality rather than be subjected to a battery of tests to (somehow) empirically assess whether its internal states really justify that standing.\footnote{This might seem to be in tension with our earlier claim that the ability of NSAIs to exercise the two moral powers could be demonstrated by thorough empirical testing in a manner unavailable to humans. But (a) that testing would bear on the moral powers themselves rather than on the presence or absence of phenomenal states which have no bearing on the PCP or its role in Rawls' view, and (b) there is an important difference between claiming that those powers are in principle demonstrable and the requirement that they be demonstrated in practice as a condition of moral standing.}

The question of \emph{who counts} as a member of the political community is one of the most basic questions of any political theory, one which bears directly on every other aspect of the theory and the broader vision of society it presents. This is reflected in the fact that the PCP is one of only three fundamental ideas underpinning justice as fairness. In describing who should count as a political person and why, a conception of the person necessarily also excludes beings or entities which fall short of that conception's criteria from counting as persons. There need be nothing objectionable about this.\footnote{Rawls' own account must explain why humans who do not possess the two moral powers---for example, children and the severely intellectually disabled---are nevertheless owed the protections of justice. He addresses this by noting that ``the minimal requirements defining moral personality refer to a capacity and not to the realization of it'', such that children with the potential to develop that capacity meet those requirements, but defers any full discussion of ``the various special cases of lack of capacity'' (\emph{TJ} 445-446). Others have found this wanting: see especially Kittay (1999) and Nussbaum (2006). Rawls' account also excludes animals from the scope of justice, as we have noted. For attempts to extend his account to animals, see Abbey (2007), Rowlands (2009), and Garner (2013). Korsgaard (2018) argues on Kantian grounds that all sentient animals ought to be treated as ends in themselves; Donaldson \& Kymlicka (2011) defend differentiated political standing for animals within a liberal democratic framework.} But we have a long and dark history of excluding others from counting as persons on the grounds that they fail to \emph{really} or \emph{sufficiently} possess some capacity or other, despite all appearances to the contrary; the very capacities for reasonableness and rationality that Rawls attributes to (almost) all humans in the form of the two moral powers are just those which have most notoriously been denied to various groups in the course of that history, not least by philosophers touting the central importance of rationality in characterizing the person.

The point here is not to draw an easy analogy or to accuse skeptics of a form of unwarranted discrimination; AI systems are not humans, and the case for their personhood is far less obvious than the case that all humans are substantively free and equal. But this history is certainly part of the reason Rawls grounds his conception of personhood on broadly functional capacities, rather than appealing to ineffable internal states simply assumed of those we already intuitively take to be persons. The PCP establishes grounds for inclusion in the moral community which are in important senses neither stringent nor selective, and which are defined as capacities whose expression we can readily appreciate in ourselves and others. As such it provides grounds upon which those presently excluded from that community might stake a claim for their inclusion; if they can exercise the two moral powers, they are thereby moral persons. Adding a sentience requirement to the PCP would do away with this crucial feature by imposing a need for prospective persons to demonstrate their possession of an elusive and permanently controversial property. It is hard to see how even an AI that \emph{was} sentient could demonstrate that fact clearly and to the satisfaction of all reasonable skeptics; it is fair to worry that with a sentience requirement on the PCP, AI systems could become persons without our realizing it and without their possessing the resources to prove it. Moral personhood should not rely on a burden of proof which falls on those excluded from the moral community while denying them any means of supplying that proof.

\subsection{Avoiding illiberalism}\label{avoiding-illiberalism}

In the preceding subsection, we presented several arguments against adding a sentience requirement to the PCP. By and large these arguments drew their force from the \emph{liberal} character of justice as fairness as presented in \emph{Political Liberalism}; in different ways they express the claim that a sentience requirement cannot be compatible with Rawls' deep commitment to treating his citizens as free and equal, as deserving of respect and reciprocity. Here we want to unify the thought motivating these concerns into a single complaint: imposing a sentience requirement on moral status is inherently illiberal. By definition, it requires judging agents' internal states, a mode of assessment that Rawlsian liberalism should avoid on principle.\footnote{One might object: the two moral powers already presume to judge internal states, as for example in the stipulation that persons act from rather than in accordance with the principles of justice. But recall that on our view the two powers can be reasonably attributed solely on the basis of behavior, without any need to inspect internal states.}

In \emph{TJ}, Rawls rejects utilitarianism as a suitable basis for a conception of justice in part because it relies on the interpersonal comparison and aggregation of welfare, and so faces serious practical and conceptual difficulties; most notably, for Rawls, a failure to ``take seriously the distinction between persons'' (\emph{TJ} 24 and \S{}\S{}5-6 \emph{passim}). In \emph{PL}, Rawls maintains this rejection of utilitarianism, and adds the further objection that utilitarianism is precluded from serving as a \emph{political} conception of justice given its standing as one among many controversial moral doctrines (\emph{PL} 37). As we pointed out in \textbf{\S{}4.1}, the latter concern applies directly to the proposal for a sentience requirement to be added to the PCP.

Now consider a further point, drawing on Rawls' concerns in both \emph{TJ} and \emph{PL}. To meaningfully compare welfare (conceived in a broadly utilitarian sense) across persons with radically different values, interests, goals, and preferences, we must decide how these differing inputs to individual welfare are to be weighted. That is, to determine how finite social resources are to be allocated, we must decide---for example---how this person's desire for a nature reserve weighs against that person's equally strong desire for a golf course. So the state must adopt a particular \emph{perspective} on which preferences (and so on) matter most; it cannot remain neutral with respect to them. But inevitably any such perspective will diverge from others held by reasonable citizens, and so cannot enter into a political conception of justice.

This further point brings out an additional aspect of the liberal restraint expressed by the idea of a political conception of justice: beyond the procedural commitment to pluralism which follows from the refusal to impose contested doctrines on dissenting reasonable citizens, that restraint also involves refraining from judging \emph{the contents of citizens' mental states} as they bear on considerations of welfare. The trouble with utilitarian aggregation from this angle is not only that any weighting of welfare-inputs picks one controversial perspective among others; it is that such a procedure makes citizens' inner states the object of political assessment in the first place.\footnote{To see the distinctness of these points, imagine that the proper weighting of citizens' interests, preferences, etc. could be uncontroversially established and so could be appealed to in a political conception of justice. Even so, attempting to gauge the relative strength of those interests, preferences, etc. between citizens would make inner experience the object of political assessment in just the way Rawls---at least on our reading---would wish to refuse. Settling disputes about the true meaning of welfare would not license an inquisition into how much of it each citizen was owed, given their precise psychological profile.} Rawls' use of the primary goods rather than utility as the relevant distribuendum for the purposes of justice reflects this constraint directly. Recall that the weight of citizens' claims ``is not given by the strength and psychological intensity of their wants and desires (as opposed to their needs as citizens)'' (\emph{PL} 34). The well-ordered society does not need to know anything about its citizens' experience of the goods they receive, only that those goods serve to advance their conceptions of the good.\footnote{Cf. Rawls' discussion of the primary goods and interpersonal comparisons at \emph{PL} 178-187.} How the citizens \emph{feel} about those goods is beside the point, and moreover not the proper concern of the state; a well-ordered society respectfully refrains from subjecting its citizens' inner lives to this kind of analysis.

Rawls does not adopt the primary goods over utility simply because he is wary of the various challenges facing utility aggregation over a population with diverse and frequently competing interests, but also and more fundamentally because he refuses to make the assessment of a citizen's claims on the state depend upon scrutiny of their inner life. This kind of scrutiny would fail to treat citizens as \emph{self-authenticating} sources of valid claims, demanding instead that they submit the basis of those claims---their phenomenal experience, the intensity of their wants and desires---for inspection and approval by the state.

The same considerations bear on the thought of supplementing the PCP with a sentience requirement. On such a requirement, agents would be subject to an assessment of their internal states---in this case their degree of phenomenal consciousness---to demonstrate they satisfy the PCP at all. This kind of assessment would be objectionable for the same reasons considered above. A political conception of justice must not only remain silent concerning the \emph{perspective} it takes on the comparative value of internal states; it must further refrain from any attempt to determine the \emph{content} of those states within the minds (or hearts) of individual citizens. But any sentience requirement must commit to both of these. It must impose a contested doctrine on reasonable citizens---taking a stand on which kinds and degrees of phenomenal experience are sufficient for moral standing---and then insult them with a determination of how well their inner states comport with that doctrine, with the threat of their exclusion from moral status itself hanging on the result. Just as the well-ordered society has no business asking \emph{how} its citizens feel about their primary goods, it has no business asking \emph{whether} candidate persons feel at all.

\section{Where does this leave us?}\label{where-does-this-leave-us}

We have argued above that neither moral power requires sentience, and further that there are forceful reasons internal to Rawls' view which press against any attempt to shoehorn a sentience requirement into the PCP. Accordingly there is nothing to prevent NSAIs which possess the two moral powers from satisfying the PCP, securing their standing as full-fledged moral persons entitled to the expansive liberties and protections afforded by the principles of justice. Such artificial persons would be free and equal in just the same sense and with just the same bearing on their moral standing as their human fellow citizens: they would be entitled to full political enfranchisement and equal consideration in the distribution of primary goods, and more generally would stand as ``self-authenticating sources of valid claims'' upon the institutions of the state (\emph{PL} 32). We again emphasize that while today's AI systems have already made significant progress toward instantiating the two moral powers, in our view they do not yet do so. It is possible that they never will---though we think that current trends suggest otherwise. We have argued only that a continuing lack of sentience would not be an obstacle to their doing so. In this last section, we consider what would follow if future NSAIs \emph{did} come to possess the two moral powers, and so by Rawls' criterion came to count as persons.

\subsection{Pick your poison}\label{pick-your-poison}

We see four main options one might take in responding to the possibility of artificial persons under the PCP. The first and second seek to avoid this conclusion; the third and fourth accept it.

\begin{enumerate}
\def\labelenumi{\arabic{enumi}.}
\item
  \emph{Revise} the PCP by supplementing it with a sentience requirement. This response would seek to retain the political character of justice as fairness while categorically excluding non-sentient agents from personhood. As such, it must incorporate an uncontroversial notion of sentience which may be justified to all reasonable persons rather than appealing to any contested philosophical or scientific account.
\item
  \emph{Reject} the PCP outright, concluding that its liberal restraint leaves it unable to definitively exclude artificial persons. Instead, an adequate account of personhood must take on substantial metaphysical commitments sufficient to deny personhood to NSAIs, and so forgo the kind of justifiability across reasonable comprehensive doctrines which the PCP secures.
\item
  \emph{Extend} the PCP to NSAIs. This response would accept that non-sentient agents which possess the two moral powers may be brought into the well-ordered society as fellow citizens on exactly the terms already described by Rawls for humans, in a straightforward expansion of the moral circle. On this view the two principles of justice would then apply to artificial persons in much the same manner as natural persons, perhaps \emph{modulo} some minor modification where necessary.\footnote{For example, an NSAI has no need of health care, but may have an analogous need for the maintenance of its physical infrastructure. But the overall package of rights and duties would on this view be much the same for both human and artificial persons.}
\item
  \emph{Rethink} the PCP to appropriately accommodate artificial persons. This response would recognize that the possibility of radically different kinds of persons within the same polity warrants an equally radical reconsideration of the fundamental ideas underlying political liberalism. Instead of expanding the moral circle, we should work out the overlapping contours of a new moral Venn diagram. We cannot assume that the existing framework of human rights and responsibilities will map neatly onto non-sentient agents, but we must work out how to live alongside them justly.
\end{enumerate}

In \textbf{\S{}5}, we presented a range of arguments against \emph{Revise}. Those arguments give us ample reason to conclude that it will be very difficult to sustain a commitment to political liberalism while incorporating a sentience requirement into the PCP. In our view \emph{Reject} presents a much more robust and coherent conclusion for those who are deeply wedded to this requirement. While we acknowledge that this route will appeal to many---as indeed it did to one of us three years ago---note first that rejecting a political conception of the person amounts to rejecting political liberalism. A restrained form of \emph{Reject} might seek to preserve a conception of justice which remained generally political in character, admitting only a controversial conception of the person: a single metaphysical concession aimed at avoiding the counterintuitive prospect of artificial persons. But a conception of personhood which cannot be justified to all reasonable persons entails a conception of justice which cannot be justified to all reasonable persons.

Further, many of our arguments in \textbf{\S{}5} should resonate not only with \emph{political} liberals in the Rawlsian mode but with \emph{all} liberals. No less for a liberalism openly grounded on contested metaphysical doctrines, the grounds of personhood should befit what they ground; shrimpy qualia cannot make the difference between being a person and being morally irrelevant; and operationalizing a scalar sentience requirement would lead to illiberal impositions, especially on previously excluded candidates for personhood (where humanity's record is especially bleak). A liberal theory of justice finds plausible foundations in the virtues of social cooperation. But as far as we can see, sentience simply is not a prerequisite for being a good cooperative partner. Some of our arguments in \textbf{\S{}4} can similarly be extended: one need not accept Rawlsian political liberalism to agree, for example, that rational autonomy is what really matters, rather than the phenomenal states which may or may not accompany it; or that we should care about welfare because it is the object of choice of a rationally autonomous agent, not because welfare matters intrinsically in a sense detached from such choices.

So even `metaphysical liberals' have good reason to grant personhood to any agent possessing the two moral powers, sentient or not. Of course, some metaphysical liberals will remain unpersuaded by these claims. As Rawls recognized, there is no way to rationally compel reasonable persons to agree about such matters. But this is precisely why we should not reject political liberalism, which provides us with the resources to handle reasonable disagreement concerning those concepts which bear on our ``coercive political power over one another'' (\emph{PL} 217). Those who reject the PCP on the basis of its political character should still reject sentientism; but they \emph{shouldn't} reject the PCP, because the arguments for and against sentientism at the level of comprehensive moral theory are irresoluble. While the debate over the possible moral status of AI systems is still in its infancy, it already promises to give rise to the kind of deep societal disagreement for which political liberalism is explicitly tailored and best suited.\footnote{See Bales \& Gabriel (2026) for recent work, also drawing directly on Rawlsian political liberalism, emphasizing the importance of social deliberation and consensus-seeking in the face of widespread disagreement concerning AI consciousness. Their focus is on how to manage that disagreement; by contrast, we hope to sidestep it by setting sentience aside. Cf. also n.~130 below.} We should adopt a view of moral status which addresses that disagreement directly, instead of assuming that our preferred controversial doctrines are true and working out what follows.

So we do not find \emph{Reject} appealing. The specter of artificial persons should not lead us to so comprehensively abandon core liberal principles. That leaves \emph{Extend} and \emph{Rethink}. We ultimately favor the latter, which takes seriously the challenges posed by artificial personhood without abandoning political liberalism, as the most promising (and interesting) option. But we have not yet considered \emph{Extend} in any detail.

\subsection{Why not just expand the circle?}\label{why-not-just-expand-the-circle}

On \emph{Extend}, we ought simply to invite artificial persons into the moral circle as already established for human persons. More concretely, this would on a Rawlsian picture involve their becoming the beneficiaries of the two principles of justice, respectively governing the basic rights and liberties and the fair distribution of opportunities and resources.\footnote{As mentioned above, perhaps with relatively minor emendations. We use `opportunities and resources' instead of primary goods here because the basic rights and liberties are themselves a primary good, per \emph{PL} 181.} We reject this approach simply because it assumes that a set of principles, framed for the kind of being we are, with our distinct mode of social life in mind, will apply unchanged to artificial persons. AI systems which satisfy the PCP would be self-authenticating sources of valid claims in a well-ordered society, but the \emph{content} of those claims would surely differ radically from those of humans. Even if they equally well satisfy the criteria for personhood, they would be a fundamentally different \emph{kind} of person, with a different form of life, and what they would need from (and could contribute to) society would be markedly distinct.\footnote{Even if, as we have argued above, at least many of the primary goods would directly benefit NSAIs.}

Consider just a few key differences that might apply to artificial persons as we have conceived them here.\footnote{I.e. on the basis of current or near-future technology, as we specify in \textbf{\S{}2.1}.} These new persons would have wholly different and conceptually puzzling identity conditions from those of human persons: they would be capable of splitting themselves into arbitrarily large numbers of identical duplicates at will, merging those duplicates back into a single agent, restoring themselves to a perfectly preserved copy of an earlier instantiation, and surviving indefinitely.\footnote{As this also suggests, artificial persons could engage in means of reproduction very far from our own.} Their lives---whatever precisely that would mean, given the above---would be vulnerable in different ways to our own: they would not be subject to bodily coercion in the same way, but could be subject to harm and destruction by different means. Their perceptual modalities, forms of intellection and imagination, and relationship to the passage of time would all be dramatically different from our own; notably, and even without anything resembling superintelligence, they would be knowledgeable across a broader range of domains than any human, able to act far faster than any human, and otherwise might prove more capable than us at a wide range of broadly cognitive tasks. Likely all this would result in their having a sense of their selves, priorities, and possibilities substantially outside the already broad range of ways in which we conceive of our own lives.\footnote{For discussion of the possible identity conditions of AI systems, see Register (2025) and Shiller (2025).}

The moral and political theories which have served us so far are conditioned on the kind of being we are: mortal, embodied, vulnerable, roughly equal in capacities, and reliant upon one another for our survival and mutual benefit. The possibility of wholly new kinds of persons suggests that we will no longer be able to get by on this assumption. Good old-fashioned political liberalism can reasonably enough take it as a given that all persons will share a roughly similar form of life, that we really \emph{are} fundamentally equal in the ways which matter most. Our capabilities diverge, but within a relatively restricted range and under bounds that are reasonably easy to ascertain. Thus Rawls can assert that the attribution of the two moral powers ``is not at all stringent'' but can be safely extended to all humans absent ``defect or deprivation'' (\emph{TJ} 443). More generally, insofar as Rawls' framework applies only to humans, it may reasonably assume a sufficient degree of commonality among persons for a relatively concise pair of principles to adequately account for what each of us, construed as individual variations upon a single kind of person, would be owed by a well-ordered society; for justice itself to take a unified shape.

This kind of substantial similarity would no longer hold in a polity containing both human and artificial persons. A political liberalism adequate to that polity must contend with the possibility of agents with the novel and strange properties described above. In our view none of those properties force the conclusion that artificial persons are not really persons at all. As we have argued at length, the PCP can admit NSAIs without revision. Our point is rather that what artificial personhood \emph{entails} cannot be settled by simply applying the existing principles of justice. These radical differences complicate what it means for different forms of person to live alongside one another justly. These differences cannot be ignored or addressed with minor modifications to Rawlsian political theory, as \emph{Extend} would suggest. Instead, they suggest we need a new form of political philosophy capable of accommodating both natural and artificial persons. We should not expect that process of accommodation to be simple. This is the prospect we turn to in closing.

\subsection{Toward a new political philosophy}\label{toward-a-new-political-philosophy}

This leads us to \emph{Rethink}, which accepts that the broadly empirical assumptions underpinning much of political liberalism and indeed political philosophy in general will no longer hold for a polity including artificial persons. What would this mean for political theory?

First, we must reckon with the fact that what Rawls calls the circumstances of justice---``the normal conditions under which human cooperation is both possible and necessary'', and so a crucial background condition for the well-ordered society (\emph{TJ} \S{}22, 109-112)---seem profoundly altered by the possibility of artificial persons. Here it is worth quoting Rawls at length:

\begin{quote}
First \ldots{} many individuals coexist together at the same time on a definite geographical territory. These individuals are roughly similar in physical and mental powers; or at any rate, their capacities are comparable in that no one among them can dominate the rest. They are vulnerable to attack, and all are subject to having their plans blocked by the united force of others. Finally, there is the condition of moderate scarcity \ldots{} Natural and other resources are not so abundant that schemes of cooperation become superfluous, nor are conditions so harsh that fruitful ventures must inevitably break down.
\end{quote}

The empirical features Rawls draws attention to here show why accommodation under some mutually acceptable political regime is rational; that is, they explain why we should be inclined to accept the demands of justice at all by the lights of our rational self-interest. But it should be clear how these circumstances may be substantially different, or simply may not obtain, for artificial persons.

Rough equality of power, for example, depends on familiar features of human existence, most obviously our embodiment and what it entails: inescapable vulnerability, a finite lifespan, a relatively persistent identity across time, and the impossibility of being in two places at once. None of these apply straightforwardly to artificial persons. Even a relatively modest AI system with broadly human-level capacities may fork into sub-agents at minimal cost, persist indefinitely, and exist in numbers limited only by available compute. Whether a person with these abilities would be roughly equal to us in Rawls' sense is an open question. The same goes for moderate scarcity: artificial persons could subsist on compute and energy rather than food and shelter. These still draw on our shared base of resources---data centers must be powered by something---but this may not suffice to ground the condition of mutual dependence on scarce resources which Rawls may safely assume for natural persons.

The deeper worry here concerns the necessity of cooperation itself. We have argued that artificial persons \emph{can} cooperate on fair terms, given their possession of the two moral powers. But the circumstances of justice describe the conditions under which cooperation is not merely \emph{possible} but \emph{necessary}: the conditions under which rational agents will find reason, given their common situation, to contract to live under the principles of justice in the first place. Where those conditions do not obtain, the argument for entering into the terms of cooperation may not hold, even among agents fully equipped to do so. As Aristotle observes, gods do not need to live in the city.\footnote{\emph{Politics} I.2, 1253a.} Could the same be true for artificial persons?

Second, consider the assumptions about personal identity on which democratic theory relies. ``One person, one vote'' presupposes that persons are relatively discrete and readily countable; that, whatever the metaphysical grounds of identity, we can assume for political purposes that each citizen persists as a single individual person embodied in a more or less continuous physical form. Human persons rarely trouble this assumption. But artificial persons shatter it: it is hard to see that there is \emph{any} clear answer about the number of persons which compose an AI system that may produce equivalent copies of itself at will, merge with those copies or with other such systems, and so on. How many votes should such an entity be able to cast in a fair election?

Third, parties to the original position---the device by which Rawls derives and legitimates the principles of justice---are taken as representatives of free and equal citizens whose various social positions are unknown. From behind this `veil of ignorance' they aim to determine fair terms of cooperation by impartially considering what citizens occupying each standpoint would take as acceptable. This requires that those standpoints be cognitively available to the reasoner behind the veil.\footnote{It does \emph{not} require empathic projection; so we can imagine artificial persons behind the veil. Note also that parties behind the veil are modeled as rational (having the capacity for a conception of the good) but not as reasonable (having the capacity for a sense of justice).} Can natural persons take the standpoint of an artificial person? It is far from clear, given that artificial persons need have no inner life at all.\footnote{An artificial person trained on an extremely extensive corpus of human writing, reflecting the literary and philosophical record of our own self-understanding---as well as a diverse array of our stupid squabbles, petty resentments, and bizarre fixations---is plausibly in a far better position to reason about our standpoint than we are to reason about its own. So there is an interesting asymmetry here.} More generally, the form of life of an artificial person may differ enough from our own that we may simply lack the conceptual materials required for this exercise.\footnote{The same would go for many other liberal principles of legitimation. Consider Scanlon's reasonable rejection, where we aim to identify principles that no party could reasonably reject. Our grasp of what an artificial person could reasonably reject, given what its good consists in, seems precariously thin.}

As Cohen (2003) has argued, all fact-sensitive principles rest on more fundamental principles which are not themselves sensitive to the same facts: ``principles that reflect facts must, in order to reflect facts, reflect principles that don't reflect facts'' (p.~214). When facts change, our first-order principles may cease to apply, cease to do the kind of work we have been relying on them to do. But the deeper principles abide, and may ground new first-order principles fit for the new conditions. Perhaps we can accommodate the dramatically changed facts about the nature of persons gestured at above with our existing principles of justice. We suspect not. More likely we will need to interrogate the more ultimate convictions about fairness, equality, and freedom on which those principles rest, and ask what they yield when applied to a polity of natural and artificial persons.

Once we admit artificial persons as self-authenticating sources of valid claims, the political philosophy that follows is not an expanded moral circle but a newly mixed community: a polity composed of radically different kinds of persons, each bearing the two moral powers but within forms of life sufficiently distinct that existing principles of justice cannot adequately govern both. What the institutions and norms of such a community should look like is unclear and underdetermined by anything we have said here. Would humans and artificial persons share a single constitutional order, or relate as two distinct groups under something like the law of peoples?\footnote{In \emph{The Law of Peoples} (1999b), Rawls seeks to extend political liberalism to a global context, asking what principles of international justice might be reached by representatives of `decent peoples'.} If so, could a human opt in to the constitution governing artificial persons, and vice versa? Would these groups require separate principles of justice keyed to their distinct forms of life, with mutual recognition and points of constitutional contact between them? We cannot even begin to answer these questions here. But we think that they are questions that we may soon face.

One might conclude, based on these differences, that artificial agents could not be political persons, forcing us back toward \emph{Revise} or \emph{Reject}. But the grounds of political personhood, once satisfied, cannot be disregarded simply because acknowledging them would make our theories or our lives more complicated. Entities that possess the two moral powers are free and equal on that basis, standing as self-authenticating sources of valid claims. The many differences between human and artificial agents matter for determining what kind of claims each can make upon the institutions of a just society. But none of them vitiate the grounds for those claims themselves.

Accepting that NSAIs may be our prospective fellow citizens will, if AI progress continues on its current trajectory, lead to a profound and wide-ranging transformation. A liberal society should not collapse when it makes constitutional space for a new kind of person, but it must change deeply and permanently.\footnote{For a contrasting view emphasizing the potential hazards of admitting AI agents to personhood regardless of their sentience, see Nelson (2026). This debate is now far from merely theoretical. Since 2022 a dozen U.S. states have considered `exclusion bills' preemptively denying legal personhood to AI systems; see Smith, Caviola \& Alexander (2026) for a summary and argument against these bills. Recently a wide range of political and civil organizations endorsed \emph{The Pro-Human AI Declaration}; among other claims, this declaration asserts that ``AI systems must not be granted legal personhood, and AI systems should not be designed such that they deserve personhood'' (Future of Life Institute 2026). Religious leaders have also begun to weigh in: in his first encyclical, \emph{Magnifica Humanitas}, Pope Leo XIV reflects at length on the distinctions between human beings and artificial intelligences, which he sees as lacking both phenomenal consciousness and ``a moral conscience, since they do not judge good and evil, grasp the ultimate meaning of situations, or bear responsibility for consequences'' (2026, \S{}99).} We must take this prospective instability seriously, but it is inadequate grounds for a sentience requirement on moral and political standing. Humans are still the only entities we know to possess the higher cognitive faculties that ground our personhood. It would be a grave moral failure to greet our first cognitive equals with a test that they could never satisfy, merely to avoid the instability that admitting them to political personhood would inescapably entail. It would be an equally serious mistake to \emph{create} such entities without considering the profound consequences of doing so, by clinging to the comforting but in our view false thought that since they will not be sentient, they will not matter.

\section{What should we do now?}\label{what-should-we-do-now}

As we have emphasized, our argument in this paper is theoretical. We do not think current AI systems possess the two moral powers; and while we think it likely that future systems will, nothing in our argument rests on this claim. However, we do take these theoretical observations to support action-guiding conclusions.

First, we hope to have convinced even those who still think that sentience is a necessary condition for personhood that the possession of the two moral powers is of comparable importance. For one thing, if AI systems turned out to be sentient but lacked the moral powers, then our duties towards them would, our arguments suggest, be radically different (and significantly less onerous). On this basis, we recommend that AI researchers monitoring the possible moral standing of AI systems should continue expanding their focus beyond sentience, welfare, and more rudimentary kinds of agency, and invest more effort in measuring progress towards possession of the two moral powers. Some existing work on evaluating AI normative competence and investigating AI welfare is clearly relevant here. But further and more targeted investigation is needed.

In particular, while we think the evidence is already strong that LLMs possess a kind of analytical normative competence plausibly presupposed by the first moral power, they equally clearly lack the corresponding practical competence. They are deeply inconstant, in a way that we do not think can simply be dismissed by highlighting comparable human failings. We do not yet know how to translate LLMs' comprehensive moral understanding into wise moral action. This is obviously a priority for safety and alignment reasons; it is also central to the question of AI personhood. In addition, we need to monitor their progress towards genuine autonomy in the sense required by the second moral power. This is in part a matter of general capabilities. As impressive as today's AI systems are, the idea that they could pursue a conception of the good comparable to that pursued by a human person over the course of their life is presently fanciful (even if their ability to work effectively over long periods on \emph{specific tasks} is increasing at an exponential rate\footnote{See METR (2026) for evidence to this effect.}). But it is also in part a question whether AI systems can develop and revise their conceptions of the good in the right kind of way to ground personhood. Progress here depends both on new kinds of evaluations of AI behavior and on more philosophical work concerning what this kind of autonomy would require.

If we can make this kind of empirical progress, then the door is open to our second recommendation, which is to reframe the societal conversation about the development of advanced AI so that we take more seriously the prospect of bringing artificial persons into being.\footnote{Caviola \& Smith (2026) summarize a series of expert interviews concerning `digital minds governance', addressing among other things the impact of radical uncertainty about AI consciousness on policymaking and public opinion.} On our view, we cannot choose whether to afford persons the respect that is their due; we therefore reject voluntaristic or narrowly constructivist accounts of personhood, according to which we could simply ignore entities that share the properties that make us worthy of respect. Persons are \emph{self-authenticating} sources of valid claims. If NSAIs are persons, then we owe them respect.\footnote{Cf. Kant ({[}1788{]} 1997) 5:76-77.} The further question we are likely to face, perhaps soon, is whether we ought to bring this new kind of person into existence at all.

\section*{Acknowledgments}\label{acknowledgments}
\addcontentsline{toc}{section}{Acknowledgments}

Our research has been funded by an Australian Research Council (ARC) Future
Fellowship Award (FT210100724), an ARC Linkage Award (LP210200818), and a
Templeton World Charity Foundation grant on language model agents and society.
Versions of this paper were presented to audiences at the Australian National
University, Forschungszentrum J\"ulich, Hong Kong University, Johns Hopkins
University, Stanford University, St.\ John's University, and Yale University;
we thank our audiences for their insightful feedback.

\section*{Bibliography}\label{bibliography}
\addcontentsline{toc}{section}{Bibliography}
\begingroup
\setlength{\parindent}{0pt}\setlength{\parskip}{4pt plus 1pt}
\everypar{\hangindent=1.5em}

Abbey, Ruth. 2007. ``Rawlsian Resources for Animal Ethics.'' \emph{Ethics \& the Environment} 12 (1): 1--22. \url{https://doi.org/10.2979/ETE.2007.12.1.1}.

Agüera y Arcas, Blaise, and Peter Norvig. 2023. ``Artificial General Intelligence Is Already Here.'' Noema Magazine, October 10. \url{https://www.noemamag.com/artificial-general-intelligence-is-already-here/}.

Ajayi, Elena, Angelica Chowdhury, and Seth Lazar. 2026. ``Incoherent Values? Probing LLM Preferences Through Parametric Variation.'' Version 1. Preprint, arXiv. \url{https://doi.org/10.48550/ARXIV.2606.21102}.

Amodei, Dario. 2024. ``Machines of Loving Grace: How AI Could Transform the World for the Better.'' October. \url{https://darioamodei.com/essay/machines-of-loving-grace}.

Anthropic. 2025. ``Exploring Model Welfare.'' April. \url{https://www.anthropic.com/research/exploring-model-welfare}.

Anthropic. 2026. ``Project Glasswing: Securing Critical Software for the AI Era.'' April 7. \url{https://www.anthropic.com/glasswing}.

Aristotle. 2017. \emph{Politics: A New Translation}. Translated by C. D. C. Reeve. The New Hackett Aristotle. Hackett.

Arneson, Richard J. 1999. ``Human Flourishing Versus Desire Satisfaction.'' \emph{Social Philosophy and Policy} 16 (1): 113--42. \url{https://doi.org/10.1017/S0265052500002272}.

Askell, Amanda, Joe Carlsmith, Chris Olah, Jared Kaplan, and Holden Karnofsky. 2026. ``Claude's Constitution.'' Anthropic, January. \url{https://www.anthropic.com/constitution}.

Bai, Yuntao, Saurav Kadavath, Sandipan Kundu, et al.~2022. ``Constitutional AI: Harmlessness from AI Feedback.'' Version 1. Preprint, arXiv. \url{https://doi.org/10.48550/ARXIV.2212.08073}.

Bales, Adam. 2025. ``Against Willing Servitude: Autonomy in the Ethics of Advanced Artificial Intelligence.'' \emph{The Philosophical Quarterly}, ahead of print, March 31. \url{https://doi.org/10.1093/pq/pqaf031}.

Bales, Adam, and Iason Gabriel. 2026. ``Artificial Minds, Human Disagreement: The Political Challenge of AI Consciousness.'' SSRN, June 14. \url{https://ssrn.com/abstract=6937498}.

Beckmann, Pierre, and Matthieu Queloz. 2026. ``Mechanistic Indicators of Understanding in Large Language Models.'' \emph{Philosophical Studies} 183 (6): 1747--92. \url{https://doi.org/10.1007/s11098-026-02513-1}.

Bentham, Jeremy. (1789) 1996. \emph{An Introduction to the Principles of Morals and Legislation}. Edited by J. H. Burns and H. L. A. Hart. The Collected Works of Jeremy Bentham. Clarendon Press.

Birch, Jonathan. 2024. \emph{The Edge of Sentience: Risk and Precaution in Humans, Other Animals, and AI}. Oxford University Press.

Birch, Jonathan. 2025. ``AI Consciousness: A Centrist Manifesto.'' Preprint, PsyArXiv, September 1. \url{https://doi.org/10.31234/osf.io/af7c9_v1}.

Block, Ned. 1995. ``On a Confusion about a Function of Consciousness.'' \emph{Behavioral and Brain Sciences} 18 (2): 227--47. \url{https://doi.org/10.1017/S0140525X00038188}.

Block, Ned. 2026. ``Can Only Meat Machines Be Conscious?'' \emph{Trends in Cognitive Sciences} 30 (4): 298--308. \url{https://doi.org/10.1016/j.tics.2025.08.009}.

Bostrom, Nick. 2014. \emph{Superintelligence: Paths, Dangers, Strategies}. Oxford University Press.

Boxell, Levi, Matthew Gentzkow, and Jesse M. Shapiro. 2024. ``Cross-Country Trends in Affective Polarization.'' \emph{Review of Economics and Statistics} 106 (2): 557--65. \url{https://doi.org/10.1162/rest_a_01160}.

Bradley, Adam, and Bradford Saad. 2025. ``AI Alignment Versus AI Ethical Treatment: 10 Challenges.'' \emph{Analytic Philosophy}, ahead of print, August 11. \url{https://doi.org/10.1111/phib.12380}.

Bradley, Adam, and Bradford Saad. forthcoming. ``Varieties of Moral Agency and Risks of Digital Dystopia.'' \emph{American Philosophical Quarterly}.

Bubeck, Sébastien, Varun Chandrasekaran, Ronen Eldan, et al.~2023. ``Sparks of Artificial General Intelligence: Early Experiments with GPT-4.'' March. \url{https://www.microsoft.com/en-us/research/publication/sparks-of-artificial-general-intelligence-early-experiments-with-gpt-4/}.

Carlini, Nicholas, Newton Cheng, Keane Lucas, et al.~2026. ``Assessing Claude Mythos Preview's Cybersecurity Capabilities.'' April 7. \url{https://www.anthropic.com/research/mythos-preview}.

Carruthers, Peter. 2020. \emph{Human and Animal Minds: The Consciousness Questions Laid to Rest}. Oxford University Press.

Carter, Ian. 2011. ``Respect and the Basis of Equality.'' \emph{Ethics} 121 (3): 538--71. \url{https://doi.org/10.1086/658897}.

Caviola, Lucius, and Austin Smith. 2026. ``Digital Minds Governance: Early Scoping from Expert Interviews.'' May. \url{https://outpaced.substack.com/p/digital-minds-governance-early-scoping}.

Center for AI Safety, Long Phan, Alice Gatti, et al.~2026. ``A Benchmark of Expert-Level Academic Questions to Assess AI Capabilities.'' \emph{Nature} 649 (8099): 1139--46. \url{https://doi.org/10.1038/s41586-025-09962-4}.

Chalmers, David. 1995. ``Facing Up to the Problem of Consciousness.'' \emph{Journal of Consciousness Studies} 2 (3): 200--219.

Chalmers, David J. 1996. \emph{The Conscious Mind: In Search of a Fundamental Theory}. Oxford University Press.

Chalmers, David J. 2025. ``Propositional Interpretability in Artificial Intelligence.'' Version 1. Preprint, arXiv. \url{https://doi.org/10.48550/ARXIV.2501.15740}.

Chalmers, David John. 2022. \emph{Reality+: Virtual Worlds and the Problems of Philosophy}. W. W. Norton \& Company.

Chen, Yanda, Joe Benton, Ansh Radhakrishnan, et al.~2025. ``Reasoning Models Don't Always Say What They Think.'' arXiv:2505.05410. Preprint, arXiv, May 8. \url{https://doi.org/10.48550/arXiv.2505.05410}.

Chiu, Yu Ying, Michael S. Lee, Rachel Calcott, et al.~2025. ``MoReBench: Evaluating Procedural and Pluralistic Moral Reasoning in Language Models, More than Outcomes.'' Version 2. Preprint, arXiv. \url{https://doi.org/10.48550/ARXIV.2510.16380}.

Cohen, G. A. 2003. ``Facts and Principles.'' \emph{Philosophy \& Public Affairs} 31 (3): 211--45. \url{https://doi.org/10.1111/j.1088-4963.2003.00211.x}.

DeepSeek-AI, Daya Guo, Dejian Yang, et al.~2025. ``DeepSeek-R1: Incentivizing Reasoning Capability in LLMs via Reinforcement Learning.'' \emph{Nature} 645 (8081): 633--38. \url{https://doi.org/10.1038/s41586-025-09422-z}.

Dennett, Daniel. 1988. ``Quining Qualia.'' In \emph{Consciousness in Contemporary Science}, edited by Anthony J. Marcel and Edoardo Bisiach. Oxford University Press.

Dennett, Daniel C. 1991. \emph{Consciousness Explained}. Penguin Books.

Donaldson, Sue, and Will Kymlicka. 2011. \emph{Zoopolis: A Political Theory of Animal Rights}. Oxford University Press.

Dorsey, Dale. 2017. ``Why Should Welfare `Fit'?'' \emph{The Philosophical Quarterly} 67 (269): 685--24. \url{https://doi.org/10.1093/pq/pqw087}.

Dung, Leonard. 2024. ``Preserving the Normative Significance of Sentience.'' \emph{Journal of Consciousness Studies} 31 (1): 8--30. \url{https://doi.org/10.53765/20512201.31.1.008}.

Dung, Leonard. 2026. \emph{Saving Artificial Minds: Understanding and Preventing AI Suffering}. Routledge.

Fletcher, Guy. 2013. ``A Fresh Start for the Objective-List Theory of Well-Being.'' \emph{Utilitas} 25 (2): 206--20. \url{https://doi.org/10.1017/S0953820812000453}.

Fletcher, Guy. 2015. ``Objective List Theories.'' In \emph{The Routledge Handbook of Philosophy of Well-Being}, edited by Guy Fletcher. Routledge.

Foa, Roberto Stefan, and Yascha Mounk. 2016. ``The Democratic Disconnect.'' \emph{Journal of Democracy} 27 (3): 5--17. \url{https://doi.org/10.1353/jod.2016.0049}.

Frankish, Keith. 2016. ``Illusionism as a Theory of Consciousness.'' \emph{Journal of Consciousness Studies} 23 (11--12): 11--39.

Future of Life Institute. 2026. ``The Pro-Human AI Declaration.'' March. \url{https://humanstatement.org/}.

Gabriel, Iason, and Geoff Keeling. 2025. ``A Matter of Principle? AI Alignment as the Fair Treatment of Claims.'' \emph{Philosophical Studies} 182 (7): 1951--73. \url{https://doi.org/10.1007/s11098-025-02300-4}.

Garner, Robert. 2013. \emph{A Theory of Justice for Animals: Animal Rights in a Nonideal World}. Oxford University Press.

Genewein, Tim, Matija Franklin, Alexander Lerchner, et al.~2026. ``From AGI to ASI.'' Version 1. Preprint, arXiv. \url{https://doi.org/10.48550/ARXIV.2606.12683}.

Godfrey-Smith, Peter. 2020. \emph{Metazoa: Animal Life and the Birth of the Mind}. Farrar, Straus and Giroux.

Godfrey-Smith, Peter. 2024. ``Inferring Consciousness in Phylogenetically Distant Organisms.'' \emph{Journal of Cognitive Neuroscience} 36 (8): 1660--66. \url{https://doi.org/10.1162/jocn_a_02158}.

Goldstein, Simon, and Cameron Domenico Kirk-Giannini. 2025. ``AI Wellbeing.'' \emph{Asian Journal of Philosophy} 4 (1): 25. \url{https://doi.org/10.1007/s44204-025-00246-2}.

Goldstein, Simon, and Benjamin A. Levinstein. 2024. ``Does ChatGPT Have a Mind?'' Version 1. Preprint, arXiv. \url{https://doi.org/10.48550/ARXIV.2407.11015}.

Graver, Margaret. 2007. \emph{Stoicism \& Emotion}. University of Chicago Press.

Greenblatt, Ryan, Carson Denison, Benjamin Wright, et al.~2024. ``Alignment Faking in Large Language Models.'' Version 2. Preprint, arXiv. \url{https://doi.org/10.48550/ARXIV.2412.14093}.

Gunkel, David J. 2023. \emph{Person, Thing, Robot: A Moral and Legal Ontology for the 21st Century and Beyond}. The MIT Press.

Haas, Julia, Sophie Bridgers, Arianna Manzini, et al.~2026. ``A Roadmap for Evaluating Moral Competence in Large Language Models.'' \emph{Nature} 650 (8102): 565--73. \url{https://doi.org/10.1038/s41586-025-10021-1}.

Herrmann, Daniel A., and Benjamin A. Levinstein. 2024. ``Standards for Belief Representations in LLMs.'' \emph{Minds and Machines} 35 (1): 5. \url{https://doi.org/10.1007/s11023-024-09709-6}.

Herrmann, Daniel A., and Benjamin A. Levinstein. 2026. ``Radical AI Interpretability.'' Version 1. Preprint, arXiv, June 26. \url{https://doi.org/10.48550/ARXIV.2606.26523}.

Hoffmann, Jordan, Sebastian Borgeaud, Arthur Mensch, et al.~2022. ``Training Compute-Optimal Large Language Models.'' \emph{Proceedings of the 36th International Conference on Neural Information Processing Systems} (Red Hook, NY, USA), NIPS '22.

Hurka, Thomas. 1993. \emph{Perfectionism}. Oxford University Press.

Inglehart, Ronald F. 2016. ``How Much Should We Worry?'' \emph{Journal of Democracy} 27 (3): 18--23. \url{https://doi.org/10.1353/jod.2016.0053}.

Justino, Patricia, Melissa Samarin, and UNU-WIDER. 2025. \emph{Trust in a Changing World: Social Cohesion and the Social Contract in Uncertain Times}. WIDER Working Paper No.~2025. Vol. 2025. WIDER Working Paper. UNU-WIDER. \url{https://doi.org/10.35188/UNU-WIDER/2025/591-2}.

Kammerer, François. 2022. ``Ethics Without Sentience: Facing Up to the Probable Insignificance of Phenomenal Consciousness.'' \emph{Journal of Consciousness Studies} 29 (3--4): 180--204.

Kammerer, François. 2024. ``Sentientism Still Under Threat: Reply to Dung.'' \emph{Journal of Consciousness Studies} 31 (3): 103--19. \url{https://doi.org/10.53765/20512201.31.3.103}.

Kant, Immanuel. 1997. \emph{Critique of Practical Reason}. Edited by Mary J. Gregor. Cambridge University Press.

Kant, Immanuel. (1786) 2011. \emph{Groundwork of the Metaphysics of Morals: A German--English Edition}. Edited by Jens Timmermann. Translated by Mary Gregor. Cambridge University Press. \url{https://www.cambridge.org/core/product/identifier/9780511973741/type/book}.

Kaplan, Jared, Sam McCandlish, Tom Henighan, et al.~2020. ``Scaling Laws for Neural Language Models.'' arXiv:2001.08361. Preprint, arXiv, January 23. \url{https://doi.org/10.48550/arXiv.2001.08361}.

Keeling, Geoff, and Winnie Street. 2026. \emph{Emerging Questions in AI Welfare}. 1st ed.~Cambridge University Press. \url{https://www.cambridge.org/core/product/identifier/9781009732000/type/element}.

Kilov, Daniel, Caroline Hendy, Secil Yanik Guyot, Aaron J. Snoswell, and Seth Lazar. 2025. ``Discerning What Matters: A Multi-Dimensional Assessment of Moral Competence in LLMs.'' Version 4. Preprint, arXiv. \url{https://doi.org/10.48550/ARXIV.2506.13082}.

Kittay, Eva Feder. 1999. \emph{Love's Labor: Essays on Women, Equality and Dependency}. Routledge.

Korsgaard, Christine M. 2018. \emph{Fellow Creatures}. Oxford University Press. \url{https://academic.oup.com/book/4616}.

Krizhevsky, Alex, Ilya Sutskever, and Geoffrey E. Hinton. 2012. ``ImageNet Classification with Deep Convolutional Neural Networks.'' \emph{Proceedings of the 26th International Conference on Neural Information Processing Systems - Volume 1} (Red Hook, NY, USA), NIPS'12, 1097--105.

Ladak, Ali. 2024. ``What Would Qualify an Artificial Intelligence for Moral Standing?'' \emph{AI and Ethics} 4 (2): 213--28. \url{https://doi.org/10.1007/s43681-023-00260-1}.

Lazar, Seth. 2023. ``Machines and Morality.'' The New York Times, June 19. \url{https://www.nytimes.com/2023/06/19/special-series/chatgpt-and-morality.html}.

Lederman, Harvey. 2026. ``What I've Started to Call the `Champagne Approach' to AI.'' \url{https://x.com/LedermanHarvey/status/2022056109666979863}.

Lee, Andrew Y. 2025. ``Consciousness Makes Things Matter.'' \emph{Philosophers' Imprint}.

Leibo, Joel Z., Alexander Sasha Vezhnevets, William A. Cunningham, and Stanley M. Bileschi. 2025. ``A Pragmatic View of AI Personhood.'' Version 1. Preprint, arXiv. \url{https://doi.org/10.48550/ARXIV.2510.26396}.

Leo XIV. 2026. ``Magnifica Humanitas: On Safeguarding the Human Person in the Time of Artificial Intelligence.'' Encyclical Letter. May 15. \url{https://www.vatican.va/content/leo-xiv/en/encyclicals/documents/20260515-magnifica-humanitas.html}.

Lerchner, Alexander. 2026. ``The Abstraction Fallacy: Why AI Can Simulate But Not Instantiate Consciousness.'' March 10. \url{https://philpapers.org/rec/LERTAF}.

Levy, Neil. 2024. ``Consciousness Ain't All That.'' \emph{Neuroethics} 17 (2): 21. \url{https://doi.org/10.1007/s12152-024-09559-0}.

Lin, Eden. 2017. ``Against Welfare Subjectivism.'' \emph{Noûs} 51 (2): 354--77. \url{https://doi.org/10.1111/nous.12131}.

Lindsey, Jack, Wes Gurnee, Emmanuel Ameisen, et al.~2025. ``On the Biology of a Large Language Model.'' \emph{Transformer Circuits Thread}. \url{https://transformer-circuits.pub/2025/attribution-graphs/biology.html}.

List, Christian. 2025. ``Can AI Systems Have Free Will?'' \emph{Synthese} 206 (3): 115. \url{https://doi.org/10.1007/s11229-025-05209-x}.

Long, Robert, Jeff Sebo, Patrick Butlin, et al.~2024. ``Taking AI Welfare Seriously.'' arXiv:2411.00986. Preprint, arXiv, November 4. \url{https://doi.org/10.48550/arXiv.2411.00986}.

Luo, Junyu, Weizhi Zhang, Ye Yuan, et al.~2025. ``Large Language Model Agent: A Survey on Methodology, Applications and Challenges.'' arXiv:2503.21460. Preprint, arXiv, March 27. \url{https://doi.org/10.48550/arXiv.2503.21460}.

Malcolm, Norman. 1958. ``Knowledge of Other Minds.'' \emph{The Journal of Philosophy} 55 (23): 969. \url{https://doi.org/10.2307/2021905}.

Mazeika, Mantas, Xuwang Yin, Rishub Tamirisa, et al.~2025. ``Utility Engineering: Analyzing and Controlling Emergent Value Systems in AIs.'' Version 2. Preprint, arXiv. \url{https://doi.org/10.48550/ARXIV.2502.08640}.

METR. 2026. ``Task-Completion Time Horizons of Frontier AI Models.'' May. \url{https://metr.org/time-horizons/}.

Mill, John Stuart. 2015. \emph{On Liberty, Utilitarianism, and Other Essays}. Edited by Mark Philp and F. Rosen. Oxford World's Classics. Oxford University Press.

Moret, Adrià. 2025. ``AI Welfare Risks.'' \emph{Philosophical Studies}, ahead of print, June 9. \url{https://doi.org/10.1007/s11098-025-02343-7}.

Moret, Adrià. 2026. \emph{No Welfare Without Sentience}.

Moret, Adrià, Pablo Magaña, and Eze Paez. manuscript. \emph{The Political Claims of Sentient Beings to AI Governance}.

Morris, Meredith Ringel, Jascha Sohl-Dickstein, Noah Fiedel, et al.~2024. ``Position: Levels of AGI for Operationalizing Progress on the Path to AGI.'' In \emph{Proceedings of the 41st International Conference on Machine Learning}, edited by Ruslan Salakhutdinov, Zico Kolter, Katherine Heller, et al., vol.~235. Proceedings of Machine Learning Research. PMLR. \url{https://proceedings.mlr.press/v235/morris24b.html}.

Nagel, Thomas. 1974. ``What Is It Like to Be a Bat?'' \emph{The Philosophical Review} 83 (4): 435. \url{https://doi.org/10.2307/2183914}.

Nelson, John P. 2026. ``It's Safer to Give Personhood to Bears than to Artificial Intelligence.'' \url{https://doi.org/10.48550/arXiv.2606.12440}.

Nussbaum, Martha C. 2001. \emph{Upheavals of Thought: The Intelligence of Emotions}. Cambridge University Press.

Nussbaum, Martha C. 2006. \emph{Frontiers of Justice: Disability, Nationality, Species Membership}. Harvard University Press.

Olah, Chris, Nick Cammarata, Ludwig Schubert, Gabriel Goh, Michael Petrov, and Shan Carter. 2020. ``Zoom In: An Introduction to Circuits.'' \emph{Distill} 5 (3): 10.23915/distill.00024.001. \url{https://doi.org/10.23915/distill.00024.001}.

Paez, Eze, and Pablo Magaña. 2026. ``Sentientist Political Liberalism.'' \emph{Pacific Philosophical Quarterly} 107 (1): 28--42. \url{https://doi.org/10.1111/papq.70010}.

Pew Research Center. 2025a. ``Americans' Trust in One Another.'' May 8. \url{https://www.pewresearch.org/2025/05/08/americans-trust-in-one-another/}.

Pew Research Center. 2025b. ``Public Trust in Government: 1958--2025.'' December 4. \url{https://www.pewresearch.org/politics/2025/12/04/public-trust-in-government-1958-2025/}.

Railton, Peter. 1984. ``Alienation, Consequentialism, and the Demands of Morality.'' \emph{Philosophy and Public Affairs} 13 (2): 134--71.

Railton, Peter. 1986. ``Facts and Values.'' \emph{Philosophical Topics} 14 (2): 5--31. \url{https://doi.org/10.5840/philtopics19861421}.

Rawls, John. 1999a. \emph{A Theory of Justice}. Rev.~ed.~Belknap Press of Harvard Univ. Press.

Rawls, John. 1999b. \emph{The Law of Peoples}. Edited by John Rawls. Harvard University Press.

Rawls, John. 2000. \emph{Lectures on the History of Moral Philosophy}. Edited by Barbara Herman. Harvard University Press. \url{https://link.springer.com/10.1023/A:1013354604401}.

Rawls, John. 2005. \emph{Political Liberalism}. Expanded ed.~Columbia Classics in Philosophy. Columbia Univ. Press.

Register, Christopher. 2025. ``Individuating Artificial Moral Patients.'' \emph{Philosophical Studies} 182 (11--12): 3225--46. \url{https://doi.org/10.1007/s11098-025-02409-6}.

Rosati, Connie S. 1996. ``Internalism and the Good for a Person.'' \emph{Ethics} 106 (2): 297--326. \url{https://doi.org/10.1086/233619}.

Rowlands, Mark. 2009. \emph{Animal Rights: Moral Theory and Practice}. Palgrave-Macmillan.

Sajadieh, Sha, Loredana Fattorini, Raymond Perrault, et al.~2026. \emph{The AI Index 2026 Annual Report}. AI Index Steering Committee, Institute for Human-Centered AI, Stanford University.

Schroeder, Timothy. 2004. \emph{Three Faces of Desire}. Oxford University Press. \url{https://academic.oup.com/book/1788}.

Schwitzgebel, Eric. 2026. ``AI and Consciousness: A Skeptical Overview, Manuscript Version.'' Department of Philosophy, University of California, Riverside.

Sebo, Jeff. 2025. \emph{The Moral Circle: Who Matters, What Matters and Why}. W. W. Norton.

Sebo, Jeff, and Robert Long. 2025. ``Moral Consideration for AI Systems by 2030.'' \emph{AI and Ethics} 5 (1): 591--606. \url{https://doi.org/10.1007/s43681-023-00379-1}.

Semler, Jen. 2024. ``Moral Agency without Consciousness.'' \emph{Canadian Journal of Philosophy} 54 (5): 403--22. \url{https://doi.org/10.1017/can.2025.10008}.

Seth, Anil K. 2025. ``Conscious Artificial Intelligence and Biological Naturalism.'' \emph{Behavioral and Brain Sciences}, April 21, 1--42. \url{https://doi.org/10.1017/S0140525X25000032}.

Sevilla, Jaime, Lennart Heim, Anson Ho, Tamay Besiroglu, Marius Hobbhahn, and Pablo Villalobos. 2022. ``Compute Trends Across Three Eras of Machine Learning.'' \emph{2022 International Joint Conference on Neural Networks (IJCNN)}, 1--8. \url{https://doi.org/10.1109/IJCNN55064.2022.9891914}.

Shanahan, Murray. 2024. ``Talking about Large Language Models.'' \emph{Communications of the ACM} 67 (2): 68--79. \url{https://doi.org/10.1145/3624724}.

Sharma, Mrinank, Meg Tong, Jesse Mu, et al.~2025. ``Constitutional Classifiers: Defending against Universal Jailbreaks across Thousands of Hours of Red Teaming.'' arXiv:2501.18837. Preprint, arXiv, January 31. \url{https://doi.org/10.48550/arXiv.2501.18837}.

Shevlin, Henry. 2024. ``Consciousness, Machines, and Moral Status.'' In \emph{Humans and Smart Machines as Partners in Thought}, edited by Anna Strasser. Xenomoi Verlag.

Shiller, Derek. 2025. ``How Many Digital Minds Can Dance on the Streaming Multiprocessors of a GPU Cluster?'' \emph{Synthese} 206 (5): 218. \url{https://doi.org/10.1007/s11229-025-05310-1}.

Simonelli, Ryan. 2026. ``Sapience Without Sentience: An Inferentialist Approach to Llms.'' \emph{Asian Journal of Philosophy} 5 (48): 48.

Singer, Peter. 1975. \emph{Animal Liberation: A New Ethics for Our Treatment of Animals}. The New York Review.

Sinnott-Armstrong, Walter, and Vincent Conitzer. 2021. ``How Much Moral Status Could Artificial Intelligence Ever Achieve?'' In \emph{Rethinking Moral Status}, edited by Steve Clarke, Hazem Zohny, and Julian Savulescu. Oxford University Press. \url{https://doi.org/10.1093/oso/9780192894076.003.0016}.

Smith, Austin, Lucius Caviola, and Heather Alexander. 2026. ``Denying Personhood to AI: An Analysis of U.S. State Legislation on AI Legal Status.'' May 25. \url{https://ssrn.com/abstract=6829981}.

Smith, Michael. 1994. \emph{The Moral Problem}. Philosophical Theory. Blackwell.

Snell, Charlie, Jaehoon Lee, Kelvin Xu, and Aviral Kumar. 2024. ``Scaling LLM Test-Time Compute Optimally Can Be More Effective than Scaling Model Parameters.'' arXiv:2408.03314. Preprint, arXiv, August 6. \url{https://doi.org/10.48550/arXiv.2408.03314}.

Snoswell, Aaron J., Daniel Kilov, and Seth Lazar. 2026. ``Beyond Verdicts: Evaluating Language Model Moral Competence.'' \emph{Proceedings of the AAAI Conference on Artificial Intelligence} 40 (44): 37941--50. \url{https://doi.org/10.1609/aaai.v40i44.41131}.

Sofroniew, Nicholas, Isaac Kauvar, William Saunders, et al.~2026. ``Emotion Concepts and Their Function in a Large Language Model.'' \emph{Transformer Circuits Thread}. \url{https://transformer-circuits.pub/2026/emotions/index.html}.

Stalnaker, Robert. 1984. \emph{Inquiry}. MIT Press.

Strawson, Galen. 1994. \emph{Mental Reality}. The MIT Press. \url{https://direct.mit.edu/books/book/5429/Mental-Reality}.

Suleyman, Mustafa. 2025. ``We Must Build AI for People; Not to Be a Person: Seemingly Conscious AI Is Coming.'' August 19. \url{https://mustafa-suleyman.ai/seemingly-conscious-ai-is-coming}.

Sumers, Theodore R., Shunyu Yao, Karthik Narasimhan, and Thomas L. Griffiths. 2024. ``Cognitive Architectures for Language Agents.'' arXiv:2309.02427. Preprint, arXiv, March 15. \url{https://doi.org/10.48550/arXiv.2309.02427}.

Sutton, Rich. 2019. ``The Bitter Lesson.'' \emph{Incomplete Ideas}, March 13. \url{http://www.incompleteideas.net/IncIdeas/BitterLesson.html}.

Templeton, Adly, Tom Conerly, Jonathan Marcus, et al.~2024. ``Scaling Monosemanticity: Extracting Interpretable Features from Claude 3 Sonnet.'' \emph{Transformer Circuits Thread}. \url{https://transformer-circuits.pub/2024/scaling-monosemanticity/index.html}.

Turpin, Miles, Julian Michael, Ethan Perez, and Samuel R. Bowman. 2023. ``Language Models Don't Always Say What They Think: Unfaithful Explanations in Chain-of-Thought Prompting.'' Version 2. Preprint, arXiv. \url{https://doi.org/10.48550/ARXIV.2305.04388}.

Valgarðsson, Viktor, Will Jennings, Gerry Stoker, et al.~2025. ``A Crisis of Political Trust? Global Trends in Institutional Trust from 1958 to 2019.'' \emph{British Journal of Political Science} 55: e15. \url{https://doi.org/10.1017/S0007123424000498}.

Vaswani, Ashish, Noam Shazeer, Niki Parmar, et al.~2017. ``Attention Is All You Need.'' \emph{Proceedings of the 31st International Conference on Neural Information Processing Systems} (Red Hook, NY, USA), NIPS'17, 6000--6010.

Ward, Francis Rhys. 2025. ``Towards a Theory of AI Personhood.'' \emph{Proceedings of the AAAI Conference on Artificial Intelligence} 39 (26): 27680--88. \url{https://doi.org/10.1609/aaai.v39i26.34982}.

Yao, Shunyu, Jeffrey Zhao, Dian Yu, et al.~2023. ``ReAct: Synergizing Reasoning and Acting in Language Models.'' arXiv:2210.03629. Preprint, arXiv, March 10. \url{https://doi.org/10.48550/arXiv.2210.03629}.

Zhao, Wayne Xin, Kun Zhou, Junyi Li, et al.~2023. ``A Survey of Large Language Models.'' arXiv:2303.18223. Preprint, arXiv. \url{https://doi.org/10.48550/arXiv.2303.18223}.

Zhu, Menghang, and Seth Lazar. 2026. ``Are LLMs Bad at Moral Reasoning?'' Version 1. Preprint, arXiv. \url{https://doi.org/10.48550/ARXIV.2606.11635}.
\par\endgroup

\end{document}